\documentstyle[epsf,psfig]{iopart}

\input{macros.phy}

\begin{document}

\title{\bf Fixed-boundary octagonal random tilings: 
a combinatorial approach}

\author{{\sc N. Destainville} \medskip \\
{\em Laboratoire de Physique Quantique~-- IRSAMC~-- UMR 5626} \medskip \\
{\em Universit\'e Paul Sabatier,} \medskip \\ 
{\em 118, route de Narbonne, 31062 Toulouse Cedex 04, France.}
\bigskip \\
{\sc R. Mosseri} \medskip \\
{\em Groupe de Physique du Solide, Tour 23, 5$^e$ \'etage,} \medskip \\
{\em Universit\'es Paris 7 et 6,} \medskip \\
{\em 2, place Jussieu, 75251 Paris Cedex 05, France.} 
\bigskip \\
{\sc F. Bailly} \medskip \\
{\em Laboratoire de Physique du Solide~-- CNRS,} \medskip \\
{\em 1, place Aristide Briand, 92195 Meudon Cedex, France.}}

\begin{abstract}

Some combinatorial properties of fixed 
boundary rhombus random tilings with octagonal symmetry are studied. A
geometrical analysis of their configuration space is given as well as
a description in terms of discrete dynamical systems, thus
generalizing previous results on the more restricted class of
codimension-one tilings. In particular this method gives access to
counting formulas, which are directly related to questions of entropy
in these statistical systems.  Methods and tools from the field of
enumerative combinatorics are used.

\end{abstract}


\section*{Introduction}

The experimental discovery of quasicrystalline
alloys~\cite{Shechtman84} led to extensive work on space tilings
over the last 15 years, as it became clear that quasiperiodic, and not
only periodic, structures could play important role in solid state
physics.  Indeed, the atomic structure of the highest quality
quasicrystals has been found to follow closely the 3-dimensional
icosahedral analogues of the celebrated pentagonal Penrose
tilings~\cite{Penrose74}.  Among the many questions that are still
open in this field, the origin of their stability is one of the
mostly highly debated.  Physical explanations range from an electronic
stabilization mechanism (refinements on the old Hume-Rothery approach)
to an original entropic stabilization, allowed by specific phason
modes which can be generated in quasiperiodic tilings. Our purpose
here is not to discuss the relative merits of the different
mechanisms, but to analyze in detail the combinatorial problems
associated with configurational entropy in random tilings.

This paper follows a previous one~\cite{Bibi97} in which the general
framework was introduced, as well as specific results concerning
codimension-one tilings. The $d$-dimensional random tilings of
interest are made of rhombi ($d$=2) or rhombohedra ($d$=3), or even
higher dimensional analogues. These tilings are projections onto a
$d$-dimensional Euclidean space of a $d$-dimensional faceted membrane
cut into a $D$-dimensional hypercubic lattice ($D>d$). The
``codimension'' of a tiling is the difference $D-d$, and the the
tiling is said to be of type $D \rightarrow d$. In
reference~\cite{Bibi97}, we discussed the codimension-one case for
tilings with specific (fixed) boundary conditions. This allows us to
write a one-to-one correspondence between tilings and combinatorial
objects, called partitions. We built a geometrical description of the
partition configuration space in terms of integral points in a high
dimensional space, the entropy being computed from the integral volume
of a specific convex polytope in that space. The occurrence of
multiplicative and additive formulas for this volume was analyzed in
detail, and given a simple geometrical meaning in the latter case in
terms of a simplicial decomposition of the convex polytope. 

The aim of the present paper is the analysis of random tilings of
higher codimension, starting with the simplest $4\rightarrow 2$
case. Studying these cases is of direct importance in the context of
quasicrystal physics, since all the quasiperiodic tilings encountered
in this field are of codimension greater than one ($5 \rightarrow 2$
for the pentagonal Penrose tiling, and $6 \rightarrow 3$ in the
icosahedral case). Tilings of type $4 \rightarrow 2$ correspond to the
so-called octagonal family, which was also observed in concrete
alloys~\cite{Kuo}. Although they are the simplest, ``octagonal''
random tilings already present most of the difficulties which, up to
now, have forbidden the derivation of exact results for the large
class of random tilings derived from hypercubic
tilings~\cite{Li92}. Note that exact results exist (for the entropy)
for other kinds of tilings, such as the square-triangle
tiling~\cite{Widom93,Kalugin94}, rectangle-triangle
tilings~\cite{Nienhuis96,Nienhuis98}, or large codimension
tilings~\cite{Widom98}. Note however that the present point of view
does not apply to the two first examples since there exist no
partition representation for such tilings.

Our analysis for the $4\rightarrow 2$ tilings follow from a
generalization of the simple partition problem, valid in case of
codimension one, to an iterated partition problem, which was proposed
earlier~\cite{Mosseri93B}, and has already led to some preliminary
numerical results. Here we describe the intricacy of the configuration
space, which is no longer convex ``as a whole'', but remains convex by
parts. We show that despite its complexity, some exact but partial
enumerative results can be obtained, although we must stress that the
ultimate goal~-- an exact formula for the entropy~-- was not obtained
and seems out of reach for the moment. We nevertheless believe that
the present analysis is an important step in at least two directions:
we give a very precise description of the configuration space and its
simplex decomposition and we point out several very closely related
problems in combinatorics, like the enumeration of sorting algorithms
(\ref{somme.des.aj}).

The paper is organized as follows. Section 1 recalls some older
results and definitions, in particular the concept of de Bruijn lines
and faceted membranes, and the bijection between standard partitions
and codimension-one tilings. Section 2 focuses on higher codimension
tilings, by introducing ``generalized'' partitions, and describing the
particular structure that is inherited by the configuration space. Its
properties in terms of local rearrangements of tiles (flips) are
analyzed in detail. In section 3, we discuss the decomposition of the
configuration space into normal simplices, and we show the latter can
be characterized thanks to a ``descent theorem''.  This allows us to
compute new enumerative formulas which were inaccessible by
``brute-force'' methods; these formulas are displayed in section 4.

Even though this paper focuses on two-dimensional tilings and more
precisely on octagonal ones, some results can easily be generalized to
higher dimensional systems. The state of the art in the $D \ra d$
cases is briefly discussed in~\ref{cycles}.

\section{Definitions and known results}

In this paper we consider 2-dimensional tilings of
rhombic tiles which fill a region of the Euclidean space without gaps
or overlaps.  The standard method for generating such structures
consists of a selection of sites and tiles in a 4-dimensional lattice
according to certain rules, followed by a projection onto the
2-dimensional subspace along a generic direction. We then say that we
have a $4 \ra 2$ tiling problem, or an octagonal one, in reference to
the sub-class of ideal quasiperiodic Ammann tilings which have
octagonal symmetry. The above procedure is also known as the
``cut-and-project'' method \cite{Elser,Duneau,Kalugin}. By
construction, the so-obtained rhombic tiles are the projections of the
2-dimensional facets of the 4-dimensional hypercubic lattice.  There
are 6 different species of tiles, two squares and four 45 degree
rhombi. Figures~\ref{worms} and \ref{ex42} show examples. In the
cut-and-project language the difference between the higher and the
lower dimensions is called the tiling {\sl codimension}. In this case
it is equal to 2.

We first recall some definitions and results which will prove to be
useful throughout this paper. These definitions are given in a
slightly more general context than the octagonal case. The higher
dimension will be denoted by $D$ and the lower one by $d$.

\subsection{De Bruijn grids and directed membranes}
\label{Grids}

Firstly, it should be mentioned that there exist two related classes of
objects which can be put in one-to-one correspondence with random
tilings: {\sl de Bruijn grids} on the one hand, and {\sl directed 
membranes} on the other hand. 

De Bruijn grids \cite{DeBruijn81,DeBruijn86} are dual representations
of tilings which can be useful to state or prove some results
concerning tilings. There are a great number of publications dealing
with these grids in the scientific literature (for example, see
\cite{Socolar85,Gahler86}), therefore we shall not give a complete
presentation of these objects.  Instead we shall give them an
intuitive definition in the case of two-dimensional tilings.
\begin{figure}[ht]
\begin{center}
\ \psfig{figure=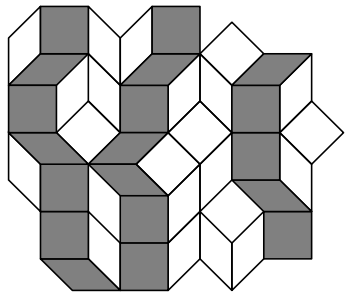,width=4.5cm} \
\end{center}
\caption{A patch of octagonal ($4 \ra
  2$) tiling. Some worms (de Bruijn lines) are represented.
  There are 4 families of worms. We have only drawn lines out of one
  of them (grayed).}
\label{worms}
\end{figure}
De Bruijn grids are made up of lines, the so-called de Bruijn lines,
which are also called ``worms''. These lines join together the middles
of opposite edges of rhombic tiles. Since the tiles are rhombi, it is
always possible to extend these lines through the tiling up to the
boundary. Such lines are displayed in figure~\ref{worms}. Any tile is
crossed by two lines. There is no triple intersection point (condition
of {\sl regularity}). On the other hand, there are lines which can
never intersect, even in an infinite tiling. They join rhombus edges
of the same orientation, as illustrated in figure~\ref{worms}. We say
that these lines belong to the same family. A family is in
correspondence with an edge orientation.  In a $D \ra 2$ tiling, there
are $D$ edge orientations and therefore $D$ families of de Bruijn
lines.

The relevant object here is not the grid itself but the underlying
intersection topology, which defines the tiling: a grid can be
directly read on a tiling by joining together the middles of opposite
edges, but it can afterwards be continuously deformed provided no
triple point appears in the process.  This grid and the tiling are
said to be {\sl dual}. In the following, we will sometimes distinguish
between the terms ``worms'', which are sequences of rhombi of a
tiling, and ``de Bruijn lines'', which are elements of a grid in an
abstract grid space, with no underlying tiling any longer.

Conversely, it can be proven that, given such a grid, it is possible to
build a unique tiling, the de Bruijn grid of which is identical to the
grid under consideration~\cite{DeBruijn81,DeBruijn86,Gahler86}.

Two lines of two different families can but need not intersect. A grid
where all lines of all families intersect is said to be {\sl
complete}. In this case, to insure the existence of all intersections,
we impose that, ``far'' from the intersection region, the lines are
perpendicular to vectors $\vect{u}_i$, one per family. The lines of a
given family are therefore parallel at the infinity.

\bigskip

Directed faceted membranes are representations of tilings in
hypercubic lattices of higher dimensions, which have been developed to
study random tilings in parallel with the partition method (see below)
\cite{Elser84,Mosseri93B,Mosseri93,Bibi97,These,Bibi98}. They are the
generalization of one-dimensional directed walks (or polymers) in
hypercubic lattices. This point of view is closely related to the
cut-and-project method. Therefore we shall only give a brief
presentation of these membranes. The main idea is that a $D \ra d$
random tiling can be lifted as a $d$-dimensional non-flat structure
embedded in a $D$-dimensional space.

\begin{figure}[ht]
\begin{center}
\ \psfig{figure=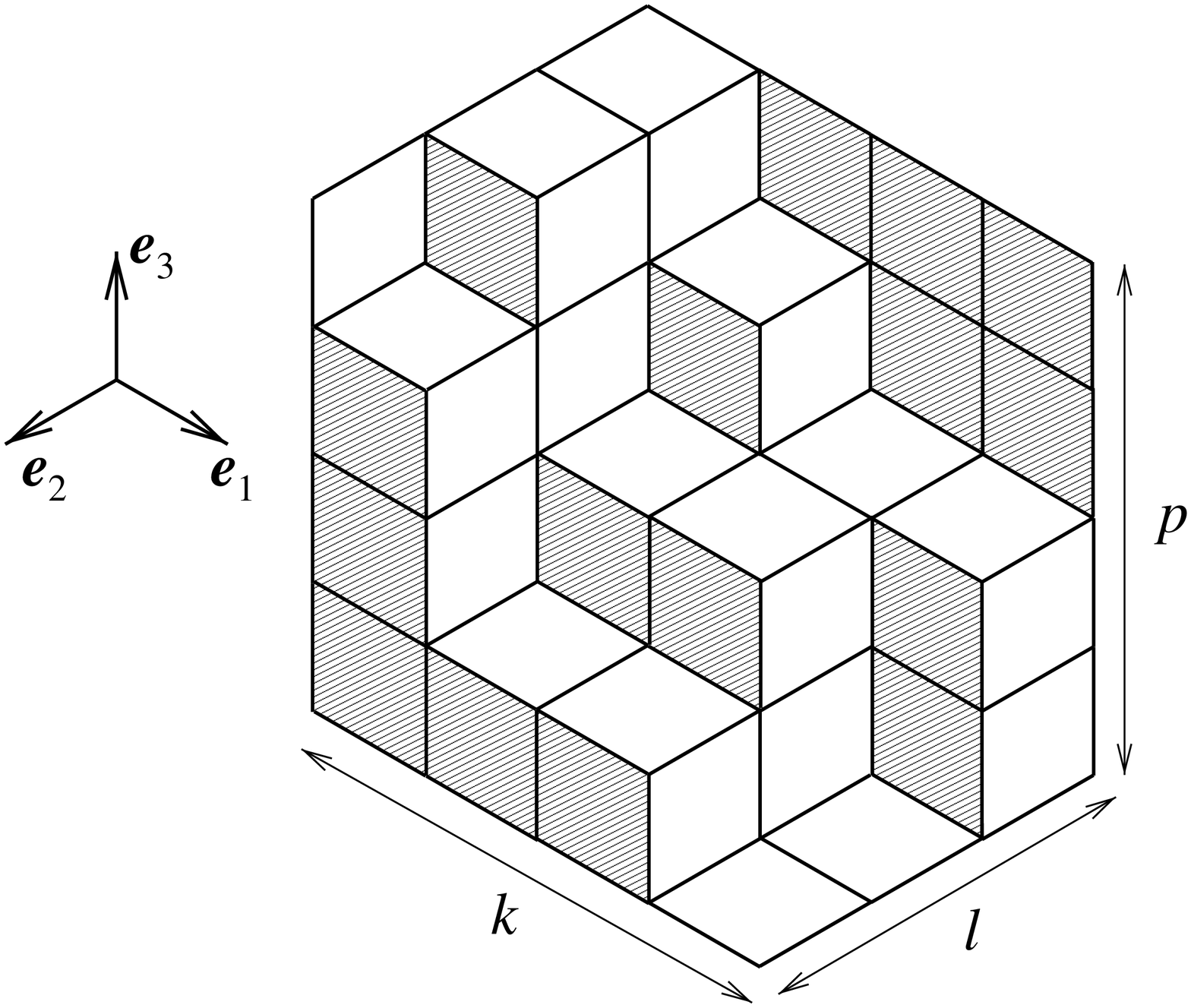,height=5cm} \
\end{center}
\caption{3-dimensional representation of a $3 \ra 2$ tiling.}
\label{ex.memb}
\end{figure}

This structure is a {\sl continuous membrane} made of $d$-dimensional
facets of the $\Zb^D$ hypercubic lattice. When this membrane is
projected along the suitable direction, the projections of these
facets are precisely the tiles the tilings are made of;
its continuous character guarantees the absence of gaps in the
so-obtained tiling. Such a membrane is said to be {\sl directed} to
emphasize the fact that its projection does not create any overlap.
For example, figure~\ref{ex.memb} displays a $3 \ra 2$ tiling, which
can also be seen as a 2-dimensional non-flat directed membrane
embedded in a cubic lattice. To get a tiling, this membrane must be
projected along the $(1,1,1)$ direction of the cubic lattice. This
point of view can be generalized to arbitrary dimensions and
codimensions. This correspondence is always one-to-one.

\subsection{Partitions in codimension one}
\label{codim1}

It is possible to derive from this membrane representation a coding of
random tilings by combinatorial objects called
``partitions''~\cite{Elser84,Mosseri93,Bibi97,These}.

This point of view is easily understood when looking at
figure~\ref{ex.memb}: the membrane can be seen as a stacking of unit
cubes in 3 dimensions and an integral height (the number of cubes) can
be assigned to each of the $kl$ columns of this stacking, resulting in
a $k \times l$ array containing integers. Since the original membrane
is directed, these numbers are decreasing in each row and in each
column of this two-dimensional array. This latter array is called a
{\sl plane partition} and each integer a {\sl part}. In this
representation, the integer $p$ is called {\sl the height} of the
partition. It is the upper bound of each part. There is a one-to-one
correspondence between such partitions and membranes embedded in a $k
\times l \times p$ piece of cubic lattice. There is a straightforward
generalization of this point of view to $D+1 \ra D$ membranes and
$D$-dimensional partitions (called {\sl hypersolid partitions}), which
are families of integers arranged in $D$-dimensional arrays,
decreasing in each direction (for more complete details, see
reference~\cite{Bibi97}, sections 2.1 and 2.2).

In the following section, we generalize this partition point of view
to any codimension tilings, which enables us to build their
configuration space. This general point of view was only briefly
tackled in previous references \cite{Mosseri93B,Mosseri93,Bibi97}. It
was developed and formalized in reference~\cite{These}.

\section{Higher codimensions tilings}
\label{GCT}

In this section, we show how it is possible to code octagonal tilings,
or more generally $D \ra d$ tilings, as {\sl generalized partitions},
that is families of integral variables, but living on structures more
complex than the previous rectangular arrays.  These structures will
turn out to be the dual graphs of relevant rhombus tilings.

\subsection{Generalized partitions}
\label{gen.parts}

Our goal in this section is to prove that $D \ra d$ tilings can also be
coded by ``generalized partitions on $(D-1) \ra d$ tilings''. Let us
explain what this terminology means.

\begin{figure}[ht]
\begin{center}
\ \psfig{figure=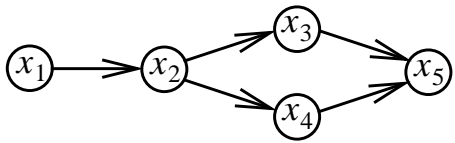,width=6cm} \
\end{center}
\vspace{-5mm}
\caption{A simple example of directed graph. It defines a partition
problem. The associated partition problem has 5 variables, which are
related by: $x_1 \geq x_2$~; $x_2 \geq x_3$~; $x_2 \geq x_4$~; $x_3
\geq x_5$~; $x_4 \geq x_5$.}
\label{ex.gen.part}
\end{figure}

Generally speaking, we define a {\sl partition problem} as a family of
$K$ integral variables, denoted by $x_1,x_2,\ldots,x_K$, placed at the
vertices of a directed graph, so that any two variables placed at two
adjacent vertices satisfy an order relation in agreement with the
orientation of the edge between those vertices\footnote{To begin with,
we shall suppose this directed graph to be {\sl acyclic}, that is to
say there is no sequence of inequalities such as $x_{i_1} \geq x_{i_2}
\geq \ldots \geq x_{i_q} \geq x_{i_1}$. The general case will be
discussed in~\ref{cycles}.}. The underlying directed graph is called
the {\sl base} of the partition problem.  To simplify, we shall
consider that all order relations are weak ($x_i \geq x_j$). The
integral values are between 0 and an integer $p$, called the {\sl
height} of the partition problem. A solution of this problem is called
a partition, of height $p$. The integral variables $x_i$ are called
the {\sl parts}. Figure~\ref{ex.gen.part} displays an example.
In reference~\cite{Bibi97}, we mainly studied hypersolid partitions,
the graph of which is equivalent to a piece a hypercubic lattice, in
the context of codimension-one partition problems (see above,
section~\ref{codim1}).

To introduce the tiling coding by partitions, we shall work in the
grid representation. We focus here on the $D \ra 2$ case (the
presentation of the general $D \ra d$ case would require some more
definitions and refinements. The interested reader will refer to
references~\cite{These,Bailey97}; see also~\ref{cycles}).  Let us
consider a $D \ra 2$ grid. We single out a family of lines, which can
be chosen as the $D$-th one without loss of generality. It contains
$k_D$ de Bruijn lines. The $D-1$ remaining families define a new
grid. We call it a {\sl subgrid} of the first one. Our goal is now to
build a partition on this subgrid that codes the initial tiling: a
part will be attached to each vertex of this subgrid.

Firstly, we need to introduce the so-called {\sl interline
indices}. Since they do not intersect, the $k_D$ singled out lines
divide the plane in $k_D+1$ domains. These domains are unambiguously
labeled from 0 to $k_D$ in the simplest way: two adjacent domains are
labeled by two successive numbers which are increasing in the
direction of $\vect{u}_D$ (as defined in section~\ref{Grids}). Now
the value of the part attached to a subgrid vertex is simply equal to
the interline index of the domain in which this subgrid vertex
lies. The maximum height of these parts is $k_D$.

There is a more simple way of characterizing the order between these
parts: since a de Bruijn line of the subgrid is transverse to all the
lines of the $D$-th family, the parts on this line are ordered in the
same direction as the interline indices. Therefore we have defined a
``canonical'' order on every subgrid line. We say that we have ordered
those de Bruijn lines. By convention, we chose those lines to be
ordered in the direction of {\sl decreasing parts} (we insist on this
point because it is a source of confusion). Now, since any two
adjacent vertices of the subgrid are joined by such a line, we have
ordered any two parts. Therefore we have defined on this subgrid a
partition problem of height $k_D$. To sum up, we have coded any $D \ra
2$ grid as a pair: a $D-1 \ra 2$ subgrid and a partition on it.

Conversely, given such a pair, the $D \ra 2$ grid from which this pair
comes can be easily re-constructed. One must add the $D$-th
family of lines in such a way that all the vertices of the subgrid lie
in the interline, the index of which is equal to the part attached to
this vertex. The constraints on the parts insure that we actually
obtain a $D \ra 2$ de Bruijn grid.

\medskip

In conclusion, we have derived a one-to-one mapping between $D \ra
2$ grids and partitions on $D-1 \ra 2$ subgrids, the parts being
suitably ordered on oriented de Bruijn lines. A
more mathematical formulation, related to this work, can be found in
reference~\cite{Bailey97}.

This mapping can be translated in the tiling (or directed membrane)
language: the generalized partitions can be defined on the suitably
oriented dual graphs of the corresponding $D-1 \ra 2$ tilings. For
short, we call them ``partitions on tilings''.

\begin{figure}[ht]
\begin{center}
\ \psfig{figure=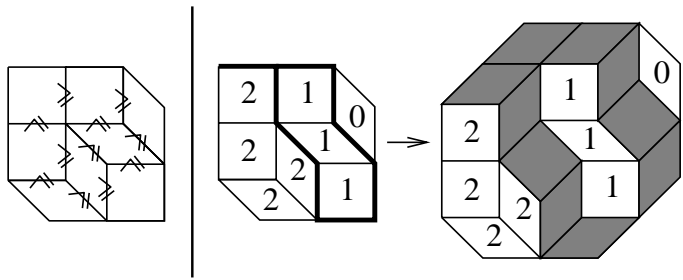,width=9cm} \
\end{center}
\caption{A $4 \ra 2$ tiling coded by a generalized partition on a $3
  \ra 2$ tiling. Left: the inequalities between the tiles. Right: a
  partition of height 2 and the corresponding $4 \ra 2$ tiling. It
  fills an octagon of sides $2,2,2,1$.}
\label{part.gen.42}
\end{figure}

Figure~\ref{part.gen.42} provides an example of $4 \ra 2$
tiling seen as a partition on a $3 \ra 2$ tiling. The $3 \ra 2$ tiling
has been slightly deformed to anticipate the next step of the process.
Note that parts are ordered on each de Bruijn line (or worm). Once the
partition has been chosen, zones where parts are equal are separated
by bold lines, which are ``opened'' to form worms (shaded) of width
1. This step is the manifestation in the tiling representation of the
fourth de Bruijn family.

To conclude this paragraph, we shall say that a $D \ra d$ tiling
problem can be studied as a collection of partition problems on a set
of (dual graphs of) $D-1 \ra d$ tilings. Even though we have only
proved this point in the $4 \ra 2$ case, the demonstration can be
generalized~\cite{These}.  Practically, to build a $D \ra d$ tiling,
one can iterate a partition-on-tiling process. The first step is
simply a codimension-one partition on a $d$-dimensional hypercubic
array. It generates a $d+1 \ra d$ tiling. The next steps increase $D$
by one each. Therefore there are $D-d$ steps.

\medskip

As compared to usual random tilings, partition-generated ones have
specific polygonal boundary conditions. For example, the tiling in
figure~\ref{ex.memb} have a hexagonal boundary. In the case of $4 \ra
2$ tilings, the polygon is an octagon of sides $k_1$, $k_2$, $k_3$
and $k_4$ (see figure~\ref{part.gen.42}).  More generally, such tilings
have {\sl zonotopal}\footnote{The zonotope generated by the
family of $d$-dimensional vectors $(\vect{v}_1,
\vect{v}_2,\ldots,\vect{v}_D)$ is the set
$$
Z = \left\{ \sum_{i=1}^D \alpha_i \vect{v}_i, \ 0 \leq \alpha_i \leq 1
\right\}.
$$
It is also called the {\sl Minkowski sum} of vectors $\vect{v}_i$. In
two dimensions, it is a $2D$-gon. The link between vectors
$\vect{v}_i$ and the $D$-dimensional representation is specified in
references \cite{Bibi97} and \cite{These}.} boundaries. Note that 
they are dual to complete de Bruijn grids. It should also be
mentioned that such polygonal boundaries have a strong macroscopic
influence on tilings, which results in a lower entropy than in free or
periodic-boundary systems ~\cite{Bibi97,Bibi98,Cohn98,Cohn9?}.

\subsection{Configuration space}
\label{config.space}

In this section, we study the configuration space of 
partition-generated tilings that fill a given polygonal domain. 

\begin{figure}[ht]
\begin{center}
\ \psfig{figure=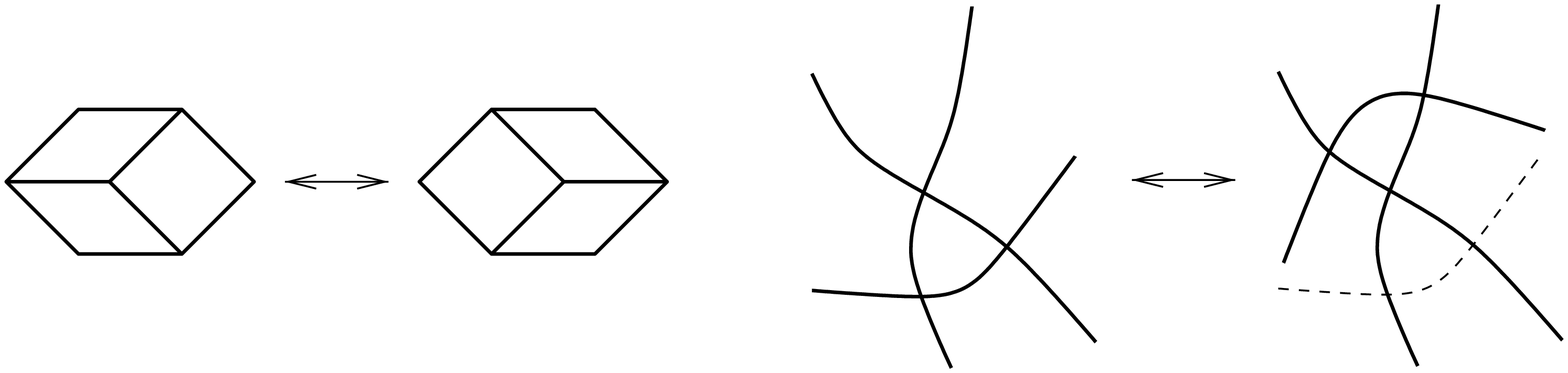,width=12cm} \
\end{center}
\caption{An elementary flip (left) and its grid space counterpart
(right).}
\label{flip}
\end{figure}

The codimension-one case has already been studied in
detail~\cite{Bibi97}: the configuration space $\CC$ consists of all
the integral coordinate points ({\sl integral points}) lying into the
convex polytope defined by the system of inequalities related to the
partition problem. This configuration space is embedded into an
Euclidean space of dimension $K$, where $K$ is the number of parts of
the partition problem. Two points are neighbors in $\CC$ (i.e. they
are linked by an edge of the underlying hypercubic lattice) if they
only differ by a local rearrangement of tiles which is usually called
an (elementary) flip~\cite{Bibi97} (see figure~\ref{flip}).

In this section, all the latter properties are extended to
generalized $4 \ra 2$ problems, in particular to partitions-on-tiling
problems. In reference \cite{These}, the general $D \ra d$ case is treated.

Let us consider $4 \ra 2$ tilings which fill an octagonal region of
sides $k_1$, $k_2$, $k_3$ and $k_4$. They are described by a class of
partition problems on $3 \ra 2$ tilings inscribed in hexagons of sides
$k_1$, $k_2$ and $k_3$. These tilings will be indexed by an integer
$\alpha$. They have exactly $K=k_1k_2 + k_1k_3 + k_2k_3$ parts. Therefore
each configuration space $\CC_\alpha$ related to the
partition problem on $\alpha$ is of dimension $K$. Now to describe the
whole configuration space of the tiling problem, we need to make explicit
how these different $\CC_\alpha$ are connected to each other.

Firstly, we need to specify how a flip in the tiling representation is
translated in the grid space. It is simply a 3-line flip, as
illustrated in figure~\ref{flip}.

If the fourth family of lines has been singled out in the
partitions-on-tiling process, two cases must be distinguished:
\begin{itemize}
\item[I:] either the 3-line flip does not involve any line from the fourth
  family. It means that the 3 vertices involved in the flip have the
  same interline index and that this index does not evolve during the
  flip. These vertices are therefore coded by parts of same value. On
  the other hand, the base tiling (i.e. the tiling dual to the
  3-family subgrid) on which the partitions are defined undergoes a
  flip; 
\item[II:] or the flip involves a line $l$ of the fourth family. In this case,
  the 3-family subgrid is not modified through the flip. The same
  holds for the base tiling on which the partitions are defined. On
  the other hand, let us consider the only vertex $S$ involved in the
  flip but which does not belong to $l$. During the flip, its
  interline index is increased by $\pm 1$. Therefore a part (and
  only one) of the partition problem varies (by $\pm 1$). 
\end{itemize}

To sum up, a flip is translated either in a base tiling flip, without
any modification of the parts (type-I flip), or in a variation of one
of the parts without any modification of the base (type-II flip).

\medskip

Let us go back to the configuration space $\CC$. Since $\CC$ can be
seen as a collection of spaces $\CC_\alpha$ associated with tilings
$\alpha$, it can be given a ``discrete fiber bundle\footnote{We employ
  improperly this term. In particular all the fibers are not
  necessarily identical.}'' structure, the base $\BC$ of which is the
configuration space of (base) tilings $\alpha$. Its fibers are the
spaces $\CC_\alpha$. 

Practically, suppose that $\BC$ is embedded in an hypercubic array of
dimension $\delta$ and that the dimension of each fiber is
$K$. Then the whole space can be embedded in a lattice $\Zb^\delta \times
\Zb^K$: the first $\delta$ coordinates code the base tilings $\alpha$ and the
$K$ last ones the parts on these tilings. We already know the
structure of $\CC$ inside a fiber: an edge between two vertices
corresponds to a type-II flip. Now, we must establish how a type-I
flip connects two fibers. 

A type-I flip consists of a flip in the base tiling, transforming the
tiling $\alpha$ into $\alpha'$, but which does not alter the values of the
parts. In the fiber, the $K$ coordinates of the corresponding points
are therefore unchanged. But in the base $\BC$, $\alpha$ and $\alpha'$
are coded by two points which differ by only one of their $\delta$
coordinates.  Thus, in $\CC$, the two tilings differ by only one
coordinate: they are neighbors in $\Zb^\delta \times \Zb^K$. 

\begin{figure}[ht]
\begin{center}
\ \psfig{figure=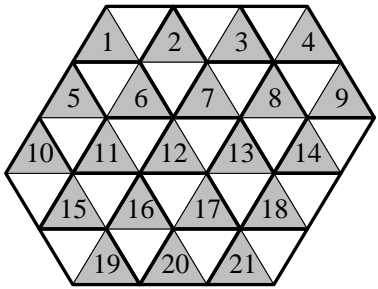,width=5cm} \
\end{center}
\caption{A $3 \ra 2$ tiling as a domino tiling on a triangular lattice. 
Each tile is labeled like its unique upward triangle.}
\label{coord.choice}
\end{figure}

However, we have omitted to deal with a subtlety in the previous
statement: so far, we have proven that two fibers are connected {\em
via} a piece of hypercubic lattice. Thus we have only proven the {\sl
local} hypercubic structure of the configuration space. To provide a
complete proof, we need to exhibit an extrinsic\footnote{That is
independent of the fiber.} set of hypercubic coordinates in which
every configuration can be encoded and in which two neighbor tilings
differ by a single flip. As a matter of fact, we only have to specify
coordinates in fibers: the choice of coordinates in the base $\BC$ is
an irrelevant question. As it was stated above, a choice of
coordinates is equivalent to the choice of a tile-labeling of a $3 \ra
2$ base tiling. Now, as illustrated in figure~\ref{coord.choice}, such
a tiling can be seen as a domino tiling on a triangular lattice: every
tile is the union of an upward and a downward triangle. Therefore
any labeling of upward triangles will provide a tile-labeling and
therefore a set of coordinates in each fiber. It is now clear that
with such coordinates, a type-I flip corresponds to a bond of the
hypercubic lattice.

\medskip

\rem we can now derive the dimension of the hypercubic lattice in
which $\CC$ is embedded:
\begin{equation}
d_\CC = \delta +K = 2 k_1k_2 + k_1k_3 + k_2k_3,
\end{equation}
since the dimension of the base is $k_1 k_2$ and the dimension of
fibers is $k_1k_2 + k_1k_3 + k_2k_3$.

\medskip

These arguments in the octagonal case can be extended by induction to the
general $D \ra d$ problem~\cite{These}, at least as far as the local
hypercubic structure is concerned (see~\ref{cycles}).  

Note also that, following this work, this configuration space has been
recently investigated further by M. Latapy, who detailed the nature of
its structure when it is seen as a partially ordered
set~\cite{Latapy99}, in any $D \ra 2$ case. Briefly speaking, the
lattice structure in the fibers looks like that of a plane partition
configuration space (it is a ``distributive lattice''), whereas the
structure of the whole set is a bit less rich (it is a lattice, 
but not a distributive one).

Another interesting question concerns the connectivity of this
configuration space: for a given boundary, is it possible to obtain a
random tiling from any other one {\em via} a sequence of elementary
flips? In the $D \ra 2$ case, the configuration space is connected for
any $D$~\cite{Kenyon93,Elnitsky97}. The present analysis provides
another straightforward proof of this result: every fiber is connected
as the configuration space of a partition problem on an acyclic
directed graph since it is the convex union of normal simplices.
Moreover, the base is connected for the same reason (inductively),
which completes the proof. The general case is discussed
in~\ref{cycles}.

\section{Decomposition of the configuration space into normal
  simplices: general case}

\subsection{Simple descent theorem}
\label{simp.DT}

We first recall the results of references~\cite{Bibi97,Stanley72}
about generalized partitions. The configuration space of a generalized
$K$-part partition problem of height $p$ is embedded in a
$K$-dimensional Euclidean space, the coordinates $x_i$ of which are
the parts of the problem.  The configurations are coded by
integral-coordinate points (called integral points), which belong to
the convex polytope $\FC^{[K]}$ defined by the intersection of the
hypercube $(0 \leq x_i \leq p)$ and of the cone $(x_i \leq x_j)$
defined by all the suitable relations between the parts.

The key point is that this configuration space can be decomposed into
elementary volumes, the so-called {\sl normal simplices}.  Let
$(\evect{1},\evect{2},\ldots,\evect{K})$ be the orthonormal basis of
the Euclidean space which generates the $\Zb^K$ lattice. A
$K$-dimensional simplex of vertices $A_0 , A_1 , \ldots , A_K$ is said
to be {\sl normal} if there exists an integer $s$ such that:
\begin{itemize}
\item each $A_i$ is an integral point,
\item $A_iA_{i+1}$ is parallel to a vector $\evect{k_i}$ for any $i$, 
\item if $i\neq j$ then $k_i \neq k_j$,
\item $\| A_iA_{i+1} \| =s$ for any $i$.
\end{itemize}
For short, we shall call such a simplex a {\em normal simplex of side
$s$}.  Its integral volume ({\em i.e.} the number of integral points
it contains) can easily be derived: it is the binomial coefficient
$\Simp{s+K}{K}$. But such simplices have lower-dimensional faces in
common that contain integral points, which must not be
double-counted. Therefore we must take into account a subtle
inclusion-exclusion scheme in order to count correctly the number of
configurations. To sum up, one must suppress $j$ faces
($j=0,\ldots,j_{\mbox{\scriptsize max}}$) to some simplices in order
to avoid double-counting. If $a_j$ is the number of simplices that
lose $j$ faces, then the number of configurations is
\begin{equation}
W(p) = \sum_{j=0}^{j_{\mathrm{max}}}
a_j \Simp{p+K-j}{K},
\label{decomp}
\end{equation}
where the maximum number of suppressed faces, $j_{\mathrm{max}}$, depends
on the partition problem under study.

The so-called {\sl descent theorem}~\cite{Bibi97} provides a
prescription to characterize the coefficients $a_j$. To state this
theorem, we need the following definition: with each simplex of the
decomposition, we can associate the sequence $(k_1,\ldots,k_K)$ of
indices appearing in the definition of the normal simplex. Then {\sl
the number of descents} in this sequence is the number of indices such
that $k_i > k_{i+1}$.

The descent theorem states that, if there exists a zero-descent
simplex, which is always true up to a re-indexing of the basis
vectors\footnote{At least in the case of an acyclic base graph
(see~\ref{cycles}).}, then a simplex with $j$ descents loses $j$
faces. Therefore the coefficient $a_j$ in equation~\ref{decomp} is
equal to the number of simplices with $j$ descents.

As a corollary, the number of normal simplices in the decomposition is
equal to the sum of the coefficients $a_j$ of equation~\ref{decomp}.

\medskip

These coefficients $a_j$ can be given a different equivalent
interpretation~\cite{Bibi97,Stanley72}: one builds a directed graph,
denoted by $T$, with two extremal vertices, $O$ and $S_0$. A
simplex of the decomposition is put in one-to-one correspondence with
{\sl maximal} walks in $T$ ({\em i.e.} going from $O$ to $S_0$). More
precisely, to each vertex of the graph, it corresponds a configuration
of height $p=1$ of the partition problem. Two configurations are
neighbors if they differ by only one part $x_i$, which is 0 in the
``lower'' configuration and 1 in the ``higher''. Therefore the link
between the two configurations can be indexed by $i$ and the descent
theorem can be translated in terms of these indices: the number of
descents of a walk in the graph is defined as the number of descents
of the sequence of indices of the bonds it follows.

To sum up, in the graph $T$ of any generalized partition problem, a
step between a vertex and one of its neighbors in a maximal walk
amounts to increasing one of the parts from 0 to 1.  Therefore a
maximal walk from $O$ to $S_0$ amounts to a labeling of the parts,
from 1 to $K$, which specifies in which order they are increased from 0
to 1. $O$ (resp. $S_0$) is the configuration where all the parts are
equal to 0 (resp. 1).

\medskip

In codimension-one $D+1 \ra D$ partition problems, we have proved that
the graph $T$ is the configuration space of the $D \ra D-1$
partition problem on a hypercubic array of sides
$k_1,k_2,\ldots,k_{D-1}$~\cite{Bibi97}.

\medskip

In codimension larger than one, that is in the case of
partition-on-tiling problems, the parts are attached to the tiles of
the $D-1 \ra d$ problem, as in figure~\ref{part.gen.42}.  Therefore,
to each maximal walk in the graph $T$, it corresponds a labelling of
the tiles, which characterizes in which order the parts are increased
by one in the walk.

For example, figure~\ref{chemin31} (left) shows a tile labeling (among
many others) in the partition problem of figure~\ref{partsur32}
(left). In fact, the only condition on those labelings is that when two
tiles $x_i$ and $x_j$ are adjacent, if the order relation is $x_i \geq
x_j$, the label associated with $x_i$ is {\sl smaller} that the
label associated with $x_j$ ($x_i$ is increased {\sl before} $x_j$).
In other words, these labels are ordered on de Bruijn lines.

\subsection{Decomposition in simplices}
\label{decomp.simp}

A $4 \ra 2$ tiling problem is a collection of generalized
partition problems on $3 \ra 2$ tilings. On each such tiling, the
descent theorem can be applied. Therefore, the counting polynomial of
the $4 \ra 2$ problem, which is the sum of all the individual
polynomials on each $3 \ra 2$ tiling, can also be written
\begin{equation}
W = \sum_{j=1}^{M} a_j \Simp{p+K-j}{K},
\end{equation}
where $K$ is the number of tiles, independent of the $3\ra 2$ tiling,
and $M$ is the greater of the integers $j_{\mathrm{max}}$ involved 
in the collection of partition-on-tiling problems.

In this section, we shall prove that the above result for
codimension-one problems~\cite{Bibi97} can be generalized: the sum of
the coefficients $a_j$ of the counting polynomial $W$ is equal to the
number of maximal walks of a given class in the related configuration
space of a given $3 \ra 1$ tiling problem. At the end of this section,
we shall give an explicit analytic expression of this number of
walks~\cite{These}. A general result concerning walks in the
configuration space can be derived in the general $D \ra d$
case. However, to avoid inessential complication, we shall present it
in the restricted $4 \ra 2$ case, and in an informal manner. A
rigorous proof in the $D \ra 2$ case is given
in~\ref{somme.des.aj.gen} and the general case is discussed
in~\ref{cycles}.

\bigskip

\begin{figure}[ht]
\begin{center}
\ \psfig{figure=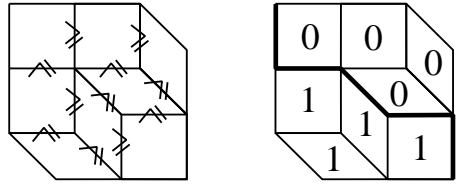,width=6cm} \
\end{center}
\caption{A partition problem on a $3 \ra 2$ tiling. Left: $\geq$
symbols show the order relations between tiles. We recall that parts
are canonically ordered on each worm.  Right: a solution of height
$p=1$ and the line separating 0's and 1's. Note that this line has
been extended up to the top-left and bottom-right corners of the
hexagon.  As a matter of fact, this line is the projection of a $3 \ra
1$ tiling.}
\label{partsur32}
\end{figure}

Let us consider a partition problem on a $3 \ra 2$ membrane, or,
equivalently, on a $3 \ra 2$ tiling. Figure~\ref{partsur32} (left) provides
an example.

\begin{figure}[ht]
\begin{center}
\begin{tabular}{cc}
        \parbox{3cm}
        {\begin{center}
        \vfill 

        \ \hspace*{-2cm}  \psfig{figure=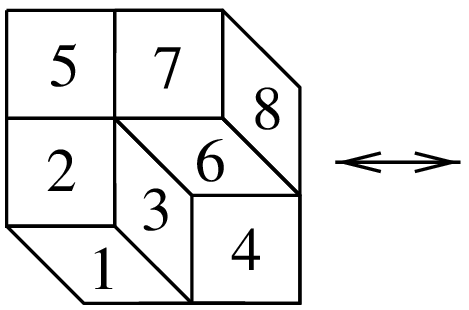,width=3cm} \

        \vfill 
        \end{center}}
        &  \parbox{2in}
        {\begin{center}
        \vfill 
        
        \ \hspace*{-1.4cm} \psfig{figure=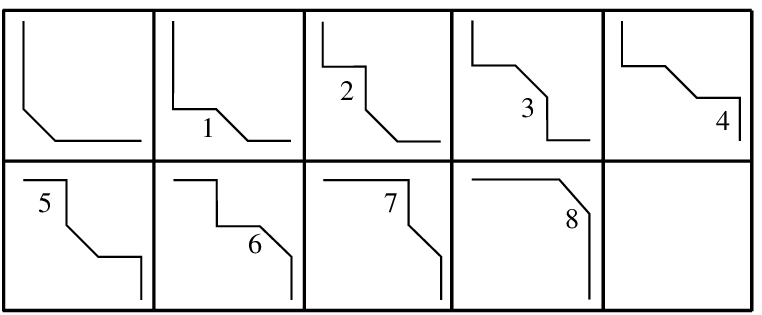,width=6cm} \

        \vfill
        \end{center}} 
        \end{tabular}
\end{center}
\caption{Left: a walk in the configuration space of the partition
  problem on the $3 \ra 2$ membrane encoded by a labeling of the 8
  tiles of this membrane. This labeling respects the order on the de
  Bruijn lines. Right: the walk in the $3 \ra 1$ problem
  configuration space associated with the left-hand-side
  labeling. One goes from a broken directed line to the next {\em via} a single
  flip. The extremal tilings are the first and the last ones.}
\label{chemin31}
\end{figure}

As we have seen it at the beginning of this section, the sum of the
coefficients $a_j$ of this (generalized) partition problem is equal to
the number of labelings of the tiles, with integers running from 1 to
$K$, which respect the following condition: these labels must be
increasing on each oriented de Bruijn line. Figure~\ref{chemin31}
(left) displays such a labeling in the case of the $3 \ra 2$ tiling of
figure~\ref{partsur32}.

On the other hand, a configuration of this partition problem of height
$p=1$ is characterized by the line which separates 0's and 1's on the
tiling. Figure~\ref{partsur32} (right) shows a partition of height 1
and the corresponding line. In figure~\ref{chemin31} (right), some
other such line configurations, associated with partitions of height
1, are displayed. These lines are directed walks using three kinds of
elementary steps: north, west and north-west ones. The number of steps
in each direction is determined by the side lengths of the hexagonal
boundary.

Now, a walk in the space of partitions of height 1 is also a walk in
the space of these directed lines. For example, the walk of
figure~\ref{chemin31} (left) is encoded into the sequence of lines of
figure~\ref{chemin31} (right). On the other hand, we notice that these
latter lines can be seen as projections in two dimensions of $3 \ra 1$
directed membranes (i.e.  one-dimensional walks embedded in a cubic
lattice). These membranes lie on the same rectangular parallelepiped
as the initial $3 \ra 2$ membrane. A walk counted by the sum of the
coefficients $a_j$ of the partition problem on this membrane is
therefore also a walk in the configuration space of a given class of
$3 \ra 1$ membranes.

To sum up, the sum of the coefficients $a_j$ of this partition problem
is equal to the number of ways of labeling the partition parts with
some rules. Such a labeling can be put in correspondence with walks
in a suitable configuration space of $3 \ra 1$ tilings. And the same
holds for each individual partition problem on a $3 \ra 2$ tiling. 

Conversely, given such a walk it is always possible to reconstruct
the $3 \ra 2$ tiling it comes from, as well as the labeling on this
tiling. Therefore this walk is counted by the sum of the coefficients
$a_j$ of a partition problem on a $3 \ra 2$ tiling, hence by the sum
of the coefficients $a_j$ of the initial $4 \ra 2$ tiling problem.

In conclusion~-- and we rigorously prove this result
in~\ref{somme.des.aj.gen}~-- the sum of the coefficients $a_j$ of a $4
\ra 2$ tiling problem is equal to the number of maximal walks in the
configuration space of a suitably related $3 \ra 1$ tiling problem.

\medskip

\noindent {\bf Theorem:} {\em For any problem of enumeration of fixed
  boundary $4 \ra 2$ tilings, the sum of the coefficients $a_j$ of the
  additive counting polynomial
\begin{equation}
W^{4 \ra 2} (p)=\sum_{j=0}^{M} a_j \Simp{p+K-j}{K}
\label{W42}
\end{equation}
is equal to the number of maximal walks in the configuration space of
the associated $3 \ra 1$ tiling problem (as defined
in~\ref{somme.des.aj.gen}).}

\medskip

Generalizing codimension-one notions, by ``maximal walk'', we mean a
walk between two tilings which play a singular role in the
configuration space, the so-called ``extremal'' tilings
(figure~\ref{chemin31}). One of the properties of these extremal
tilings is that they go in pairs and that the (Manhattan) distance
between two such configurations is a local maximum in the
configuration space, even in the general $D \ra d$
case~\cite{These}. These tilings are described
in~\ref{somme.des.aj.gen} and \ref{cycles} (see figures~\ref{ex31_32}
and \ref{macle}).

In~\ref{somme.des.aj}, we explicitly calculate this number of
walks in the $D \ra 1$ configuration spaces. In particular, if
$A_3(k,l,m)$ denotes the sum of the coefficients $a_j$ of the
polynomial counting the number of $4 \ra 2$ tilings inscribed in an
octagon of sides $(k,l,m,p)$,

\begin{eqnarray}
\label{hipps}
A_3(k,l,m) & = & (kl+km+lm)!{\left[ \fact2{(l-1)} \right]^2 
 \over \fact2{(2l+k+m-1)} } \times \\ \nonumber
 &  & \vspace{3cm} {\fact2{(k-1)} 
  \fact2{(m-1)} \fact2{(2l+k-1)} \fact2{(2l+m-1)}  \over \fact2{(2l-1)} \fact2{(l+k-1)} \fact2{(l+m-1)}}, \\ \nonumber
\end{eqnarray}
where we have used the {\sl generalized factorial functions} of order
2, as defined in reference~\cite{Mosseri93}: $\factn{2}{k} =
\displaystyle{\prod_{j=1}^k j!}$. Some properties of these functions
are given in reference~\cite{Bibi97}.

\subsection{Generalized descent theorem}
\label{descent}

As it was done in reference~\cite{Bibi97} in the codimension-one
case, we shall now refine the previous theorem in order to
characterize among the maximal walks those which are counted by a given
coefficient $a_j$. 

In other words, we are looking for a descent theorem in any
codimension. As in codimension one, we need an edge labeling in the
associated $3 \ra 1$ configuration space such that $a_j$ is the
number of maximal walks which have exactly $j$ descents with respect
to this labeling.

Now, we know that such an edge in the configuration space is in
one-to-one correspondence with a facet of a $3 \ra 2$ tiling (the
value of which changes from 0 to 1). This two-dimensional facet also
belongs to a membrane attached to the rectangular parallelepiped $P$.
Therefore it belongs to the piece of cubic lattice bounded by $P$. 

\begin{figure}[ht]
\begin{center}
\ \psfig{figure=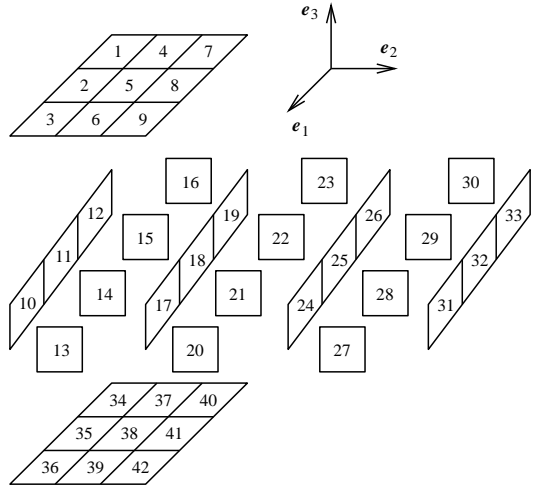,width=7cm} 
\end{center}
\caption{Labeling of the two-dimensional facets of a parallelepiped
  $P$ of sides $3 \times 3 \times 1$. For any $k_3$ larger that 1, the
  same labeling could be handled in the same way.}
\label{ordre.facettes}
\end{figure}

\begin{figure}[ht]
\begin{center}
\ \psfig{figure=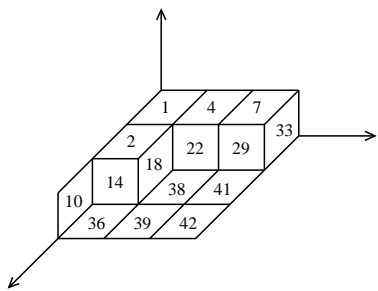,width=5cm} \
\end{center}
\caption{Example of labeling induced on a $3 \ra 2$ membrane.}
\label{ex.facettes}
\end{figure}

If we choose a labeling of all these two-dimensional facets bounded by
$P$, it will induce a labeling of the facets of any $3 \ra 2$
membrane attached to $P$. Such a labeling will be used to index the
variables of the partition problem on this membrane. For example,
figure~\ref{ordre.facettes} displays a labeling of the facets bounded
by $P$, and figure~\ref{ex.facettes} shows the induced labeling on a membrane
attached to $P$. Then, according to the descent theorem, if there
exists a zero-descent walk for this labeling, then a coefficient
$a_{j_0}$ of this individual partition problem counts the number of
$j_0$-descent walks on this membrane, and thanks to the correspondence
between facets and edges, a $j_0$-descent in the whole $3 \ra 1$
configuration space.

Conversely, let us consider any $j_0$-descent maximal walk in this
configuration space. This walk is counted by the sum of the
coefficients $a_j$ of the partition problem on an individual $3 \ra 2$
membrane. If there exists a zero-descent walk on this membrane, this
walk is more precisely counted by the coefficient $a_{j_0}$. 

In conclusion, we are looking for a labeling of the facets which
induces, on any $3 \ra 2$ membranes, a labeling with a
zero-descent walk. Then the number of $j_0$-descent walks in the $3
\ra 1$ configuration space will be equal to the coefficient $a_j$ of
the polynomial $W(p)$. We propose such a labeling in the following
paragraph.

\medskip

Let $P$ be a rectangular parallelotope of sides $k_1 \times k_2 \times
k_3$ embedded in a 3-dimensional lattice, the orthogonal basis of
which is $(\evect{1},\evect{2},\evect{3})$ (see
figure~\ref{ordre.facettes}; among all the possible choices, the
hexagonal non-flat frame of the membranes is chosen as follows: it
contains the vertices $(0,0,k_3)$ and $(k_1,k_2,0)$ of $P$).
Each facet is coded by the coordinates of its center. We define on
these coordinates an order relation $\prec$ close to the lexicographic order,
apart from a slight difference: let $f_1$ and $f_2$ be two facets, the
center coordinates of which are respectively 
$(x^1_1,x^1_2,x^1_3)$ and $(x^2_1,x^2_2,x^2_3)$. Then
\begin{itemize}
\item if $x^1_3>x^2_3$ then $f^1 \prec f^2$.
\item if $x^1_3=x^2_3$ and this value is an integer, i.e.
  these facets are horizontal\footnote{Normal to
  $\evect{3}$.}, then
\begin{itemize}
\item if $x^1_2<x^2_2$ then $f^1 \prec f^2$.
\item if $x^1_2=x^2_2$ and if $x^1_1<x^2_1$ then $f^1 \prec f^2$.
\end{itemize}
\item if $x^1_3=x^2_3$ and this value is a half-integer, i.e.
  these facets are vertical, then
\begin{itemize}
\item if $x^1_2<x^2_2$ then $f^1 \prec f^2$.
\item if $x^1_2=x^2_2$ and if $x^1_1>x^2_1$ then $f^1 \prec f^2$
(only this point differs from the lexicographic order definition).
\end{itemize}
\end{itemize}

Figure~\ref{ordre.facettes} illustrates this point. It displays a
facet labeling compatible with the previous order relation.
$P$ has been ``exploded'' to enable a better reading. Let us
emphasize the difference between horizontal and vertical layers.

In the following figure~\ref{ex.facettes}, we have represented a $3
\ra 2$ membrane attached to $P$ and its facet labeling induced by
the previous one. We check that this labeling is compatible with the
order on the de Bruijn lines (which are oriented from top to bottom
here). This result does not depend on the chosen $3 \ra 2$ membrane.

Note that the so-obtained labelings differ from those previously
defined in order to prove the hypercubic character of the
configuration space (section~\ref{config.space}).

This establishes the existence of a descent theorem in the octagonal 
case of interest. It will be used in the following section to get
finite tiling enumerations.

\medskip

\rem It is natural to wonder if there exists such a result in the
general $D \ra d$ case. Except the facet labeling, the same arguments
hold in this general context: if such a labeling existed, it would
provide a descent theorem. However, we have good reasons to believe
that such a labeling does not exist. Indeed its existence would mean
that in the general case, the counting polynomial can still be written
$\displaystyle{W(p)=\sum a_j \Simp{p+K-j}{K}}$. But we know examples
(in the $7 \ra 3$ case) where such an expression is false
(see~\ref{cycles}). Therefore if there exists a general descent
theorem, it will have a more complex statement than the previous one.

\section{Enumeration results}
\label{results}

\subsection{Two exact formulas}
In reference~\cite{Elnitsky97}, Elnitsky provides two exact formulas
for $4 \ra 2$ tiling enumeration, in the case where two sides of the
octagon are equal to 1. 

\begin{figure}[ht]
\begin{center}
\ \psfig{figure=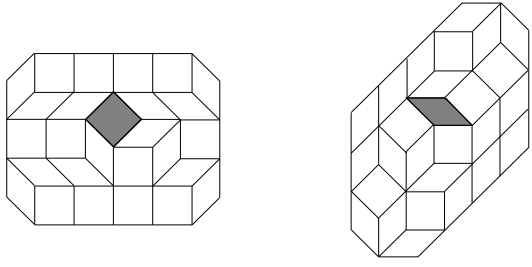,width=7cm} \
\end{center}
\caption{Two tilings of octagons of sides $r$,1,$s$,1   (left) and $r$,$s$,1,1
(right).}
\label{ex42}
\end{figure}

As displayed in figure~\ref{ex42}, two cases must be considered,
according to whether the two sides of length 1 are adjacent or
not:

\begin{equation}
W^{4 \ra 2}_{r,1,s,1}=\sum_{a+b=r} \ \sum_{c+d=s} \Simp{a+c}{a}
\Simp{b+c}{b} \Simp{a+d}{a} \Simp{b+d}{b};
\label{Elnit1}
\end{equation}

\begin{equation}
W^{4 \ra 2}_{r,s,1,1}={2 (r+s+1)! \; (r+s+2)!
\over r! \; s! \; (r+2)! \; (s+2)!}.
\label{A7}
\end{equation}
The sketches of the proofs of these formulas are recalled in~\ref{Elnitsky}.

\medskip

In his paper, Elnitsky also says that no closed formula is known for the 
first double sum. We have somewhat simplified the first formula:
\begin{equation}
W^{4 \ra 2}_{r,1,s,1} =  {(r+s+1)! \over r! \; s! \; (2r+1) (2s+1)} 
\left[ {2(r+s+1)! \over r! \; s!} 
+ \sum_{k=0}^r {1 \over 2k - 1} \Simp{r}{k} \Simp{s}{k} \right].
\label{francis}
\end{equation}
Since the last sum can be written in terms of a hypergeometric
function 
\begin{equation}
\sum_{k=0}^r {1 \over 2k - 1} \Simp{r}{k} \Simp{s}{k} = \
_3\!\, F_2 \left[ -1/2,-r,-s ; 1/2,1 ;
1 \right],
\end{equation}
this result is a ``closed'' form for the enumerating function, even
though it is not written in terms of products or ratios of simple
functions.

In order to prove equation~\ref{francis}, we need the 
following recursion relation, due to Brock~\cite{Brock}:
\begin{equation}
W^{4 \ra 2}_{r,1,s,1} - W^{4 \ra 2}_{r-1,1,s,1} - W^{4 \ra 2}_{r,1,s-1,1}
=\Simp{r+s}{r}^2.
\label{brock}
\end{equation}
The sketch of the proof of this recursion relation is also given
in~\ref{Brock}.

If the right-side member of
equation~\ref{francis} is written as the sum of two terms, we denote
the first one by $A_{r,s}$ and the second one by $B_{r,s}$. Then
\begin{equation}
A_{r,s}-A_{r-1,s}-A_{r,s-1}=\Simp{r+s}{r}^2 \left[
1 + {1 - 4rs \over (4r^2 -1)(4 s^2 -1)} \right],
\end{equation}
by simple algebraic manipulations, and
\begin{eqnarray}
 & & B_{r,s}-B_{r-1,s}-B_{r,s-1} \\ \nonumber
 & = & \Simp{r+s}{r} { 1 \over (4r^2 -1)(4 s^2 -1)} \ \ \sum_{k=0}^r 
	{1 \over 2k - 1} \Simp{r}{k} \Simp{s}{k} \times \\ \nonumber 
 &   & \ \ \left[ (2r-1)(2s-1)(r+s+1) - r(2r+1)(2s-1){r-k \over r} 
		\right. \\ \nonumber
 &   & \hspace{5.5cm} \left. -s(2s+1)(2r-1){s-k \over s} \right] \\ \nonumber
 & = & \Simp{r+s}{r} { 4rs - 1 \over (4r^2 -1)(4 s^2 -1)} \ \ \sum_{k=0}^r 
	\Simp{r}{k} \Simp{s}{k} \\ \nonumber
 & = & \Simp{r+s}{r}^2 { 4rs - 1 \over (4r^2 -1)(4 s^2 -1)}, \\ \nonumber 
\end{eqnarray}
since the last sum is equal to $\displaystyle{\Simp{r+s}{r}}$, which
proves that our expression satisfies relation~\ref{brock}. Since it
also gives the expected values for $r=0$ or $s=0$, this achieves
the proof.

\medskip

Note however that those two formulas do not give any relevant
information on the entropy since $r$ and $s$ the bigger, the most the tiling
looks like a square lattice, with two defect lines (the two worms of the
two other families) that only have a linear contribution to the entropy.
Therefore the entropy per tile vanishes when $r$ and $s$ tend to infinity.

\subsection{Numerical results}

In table~\ref{coefaj}, we list some coefficients $a_j$ obtained {\em
via} the method exposed above: we construct the oriented graph of the
configuration space of the corresponding $3 \ra 1$ problem and apply
the generalized descent theorem, as described in the codimension-one
case in reference~\cite{Bibi97}. Its label is attached to each
(oriented) edge of the graph of this configuration space, as
prescribed in the generalized descent theorem scheme, and the
coefficients $a_j$ are computed recursively: if $a_j(v)$ is the number
of walks in the configuration space from the extremal vertex $O$ to
the vertex $v$, with $j$ descents, then the sequence
$(a_j(v))_{j=0,\ldots,j_{\max}}$ only depends on the vertices under
$v$ in the graph. Then the coefficients we are interested in are equal
to $a_j(S_0)$, where $S_0$ still denotes the extremal vertex
associated with $O$.

\begin{table}
\begin{center}
\begin{tabular}{|c|l|}
\hline
$k_1,k_2,k_3$ & $a_0$ ; $a_1$ ; $\ldots$ \\ \hline \hline
2,2,2 & 20 ; 220 ; 703 ; 943 ; 566 ; 166 ; 21 ; 1 \\ \hline 
3,2,2 & 50 ; 1281 ; 9775 ; 32304 ; 53175 ; 46343 ; 22095 ; 5755 ; 774 ; 47 
; 1 \\ \hline
2,3,2 & 50 ; 1240 ; 10472 ; 40378 ; 77328 ; 75652 ; 36506 ; 7958 ; 648 ; 18
\\ \hline
2,3,3 &  175 ; 9792 ; 183223 ; 1611390 ; 7581596 ; 20313994 ; 31942744 ; \\
      & 29678550 ; 16076840 ; 4906164 ; 794328 ; 62142 ; 2088 ; 24 \\ \hline
3,2,3 &  175 ; 10372 ; 184113 ; 1445070 ; 5924665 ; 13826440 ; 19251677 ; \\ 
      & 16431348 ; 8710059 ; 2861124 ; 569191 ; 65214 ; 3943 ; 108 ; 1 
\\ \hline
3,3,3 & 980 ; 119284 ; 4736040 ; 88959048 ; 922861456 ; 5735679224 ; 
22400451966 ; \\
 & 56586512056 ; 93968296600 ; 103217016568 ; 74801020694 ; 35369632364 ; \\
 & 10693166706 ; 2003702920 ; 222619576 ; 13801976 ; 439638 ; 6272 ; 32 
\\ \hline
3,3,4 & 4116 ; 990574 ; 75291817 ; 2672974232 ; 52557540678 ; 628628119744 ; \\
 & 4845859698991 ; 25007135636872 ; 88641414434386 ; 219565301033744 ; \\
 & 384158453148998 ; 477331133707230 ; 421472964232612 ; 263431654905354 ; \\
 & 115559997005453 ; 35098071282418 ; 7238626577471 ; 987285691504 ; \\ 
 & 85977846450 ; 4564265102 ; 138792310 ; 2208928 ; 15936 ; 40
\\ \hline
3,4,4 & 24696 ; 11185183 ; 1658701257 ; 117639867825 ; 4696728888239 ; 
115554431503049 ; \\
 & 1855639954964533 ; 20237165017326054 ; 154261056214441072 ; \dots 
\\ \hline
4,4,4 & 232848 ; 211868010 ; 59911555328 ; 7889440518518 ; 578616346951691 ; \\
 & 26140019431942187 ; 775751817756005455 ; 15811577667366075305 ; \ldots
\\ \hline 
4,5,5 & 16818516 ; 52683466776 ; 49453853710872 ; 21112489152560570 ; \\
 & 4940628646460445115 ; 704860523557345706986 ; 65676322673579106872954 ; 
\ldots 
\\ \hline
5,5,5 & 267227532 ; 1658888888852 ; 2898208633474138 ; 2212967878070760376 ;\\
 & 903353585201401013350 ; 221402610595368245987868 ; \ldots
\\ \hline 
\end{tabular}
\end{center}
\caption{Coefficients $a_j$ associated with some octagonal tiling
problems.}
\label{coefaj}
\end{table}

One checks that the corresponding sums $\sum a_j$ are in agreement
with those computed in section~\ref{decomp.simp}
(equation~\ref{hipps}). Note that in table~\ref{coefaj}, the symmetry
$a_j=a_{M-j}$ observed in codimension one is lost ($M$ is the maximum
number of descents). Indeed, this symmetry remains valid for each
partition on tiling problem, but the maximum number of descents
depends on the base tiling.

The derived counting polynomials provide enumerations of tilings, as
well as entropies per tile\footnote{We recall that the entropy per
tile is $S=\ln W /N$ where $W$ is the number of configurations
(tilings) and $N$ the number of tiles.} of finite-size systems. Some
examples are listed in table~\ref{enum.ent}. The interest of the
method is that it gives access to enumeration of tilings for
arbitrarily large $k_4$, if $k_1,k_2$ and $k_3$ are fixed. Moreover it
is technically much easier to implement than a brute force enumeration
method, and very much faster as well, in terms of computational time.

\begin{table}
\begin{center}
\begin{tabular}{|c|l|l|l|}
\hline
$k_1,k_2,k_3,k_4$ & \mbox{Number of tilings} & \mbox{\# tiles}  &
\mbox{Entropy} \\ \hline \hline
1,1,1,1 & 8 & 6 & 0.34657 \\
2,2,2,2 & 5383 & 24 & 0.35796 \\
3,3,3,3 & 273976272 & 54 & 0.35979 \\
4,4,4,4 & 1043065776718923 & 96 & 0.36022 \\
5,5,5,5 & 296755610108278480324496 & 150 & 0.36031 \\
\hline
\end{tabular}
\end{center}
\caption{Some tiling enumerations computed with the previous
coefficients and the corresponding entropies per tile. We have only
listed diagonal entropies. The number of rhombi is given in
column 3.}
\label{enum.ent}
\end{table}

Even though it is not possible to make any reliable fit with few
finite-size values, it is rather clear from the available data that
the diagonal\footnote{The entropy is said to be diagonal when all the
boundary side lengths $k_i$ are equal.} entropy converges rapidly to
its limiting value. Note that in the $2 \ra 1$ case as well as in the
$3 \ra 2$ one, where exact enumeration formulas are known (see
reference~\cite{Bibi97} for a review, for instance), the asymptotic
behavior of the finite-size corrections to the entropy can be
derived~: they decrease like $\log(k)/k$. Fitting such a behavior with
the numerical values, we get a limiting diagonal entropy close to $S =
0.36(1)$. The precision of this entropy cannot be refined beyond the
second digit with the small amount of values we have got.

Previous entropy calculations {\em via} transfer-matrix
methods where derived concerning octagonal tilings, but in the case of
periodic or free boundary conditions~\cite{Li92}, leading to a
limiting value $S = 0.434$. The difference between both
results is due to the strong macroscopic effects of boundary
conditions in those random tiling systems~\cite{Elser84,Bibi98}.

\section*{Conclusion}

Among the large class of random tilings, this paper is devoted to
fixed boundary codimension-two tilings of rhombi (``octagonal
tilings''). We have established combinatorial properties of the
configurational spaces of such tilings, extending results previously
derived in the more restricted case of codimension-one tilings.
Octagonal tilings are more closely related to real quasicrystals than
are codimension-one tilings. Moreover, many of the results presented
here can (at least partially) be extended to two-dimensional tilings
of rhombi of any codimension, and even to any-dimensional tilings.

The present analysis provides additive formulas which simplify
significantly the enumeration of finite-size tilings. In a geometrical
viewpoint, these formulas come from a decomposition of the
configuration space into elementary volumes, called normal
simplices. The number of configurations in each of these simplices
is known. But it is necessary to take into account interfaces
between those volumes to avoid multiple counting, which is
achieved by the generalized descent theorem. The number of 
simplices in this decomposition is also derived in the general
two-dimensional case: these simplices are put in one-to-one
correspondence with a class of paths in a configuration space 
of tilings of same codimension, but smaller dimension, which
can be counted.

The new insight on the sets of octagonal tilings provided by this
analysis will be useful to study topics such as diffusion in these
configuration spaces, which is directly related to the rate of
convergence of flip dynamics towards the equilibrium distribution.
This problem has already been treated in the case of hexagonal
tilings~\cite{Henley97,Randall98,Randall99,Wilson99}, but is still an
open question in higher codimension plane tilings. Significant
progress will be published separately. It would also be of high
interest to understand how the introduction of energetic interactions
between tiles is translated in the configuration space and how it
modifies the dynamics. Indeed, a realistic model of quasicrystals
requires one takes into account energy, which can be in first
approximation modeled by tile interactions; glass-like slow dynamics
are likely to appear in this case~\cite{Mosseri95,Leuzzi99}, even
though no glassy behavior has been explicitely exhibited in
rhombus tilings yet~\cite{Strandburg90,Ishii95}.

\section*{Acknowledgments}

We wish to express our gratitude to Mike Widom, Matthieu Latapy and
Vic Reiner for fruitful discussions and helpful comments.

\appendix

\section{Coefficients $a_j$ and maximal walks in configuration spaces}
\label{somme.des.aj.gen}

We consider $D \ra 2$ fixed boundary tilings, described as generalized
partitions of height $k_D$ on $D-1 \ra 2$ membranes or tilings. To 
each such membrane is attached a partition problem, and therefore a 
set of coefficients $a_j$. The sum of the coefficients $a_j$ of
the counting polynomial of all these $D \ra 2$ tilings is equal
to the sum, running over all the relevant $D-1 \ra 2$ membranes,
of the sums of the coefficients $a_j$ on each such membrane.

Consider first a $D-1 \ra 2$ membrane, denoted by $\MC$, and a
partition problem on $\MC$. According to results of
section~\ref{simp.DT}, the sum of the coefficients $a_j$ of this
partition problem is equal to the number of walks, in the
configuration space of the partition problem of height 1, between two
{\sl extremal} $D-1 \ra d$ configurations, $O$ and $S_0$. All the
parts are of $O$ and $S_0$ are respectively equal to 0 and 1. Such a
walk is denoted by a sequence $O=P_0,P_1,\ldots,P_{K-1},P_K=S_0$.

We use again the grid representation of tilings and the formalism
introduced in section~\ref{gen.parts}: the partition problem on $\MC$
is seen as a partition problem on the vertices of the corresponding
grid (which has $D-1$ families), $\GC$, called the subgrid of the
problem.  The vertices of $\GC$ are ordered on each de Bruijn
family. A partition $P_i$ of height $p=1$ consists of marking each
vertex of $\GC$ by a 0 or a 1. The vertices marked with a 0 and those
marked with a 1 are separated by the only de Bruijn line of the $D$-th
family of the original grid. This latter line is denoted by
$\SC_D$. Our goal is now to encode $\SC_D$ by a $D-1 \ra 1$ grid, or
in other words by a $D - 1 \ra 1$ tiling.

Now the section of the subgrid $\GC$ by $\SC_D$ is precisely of this
type. Indeed, if we identify the intersection of a de Bruijn line 
of the $k_i$-th family and $\SC_D$ with a $i$-tile, this latter
section is a sequence of tiles, $k_i$ of each family $i$, that is to
say a $D - 1 \ra 1$ tiling.

\begin{figure}[ht]
\begin{center}
\ \parbox{8cm}{\psfig{figure=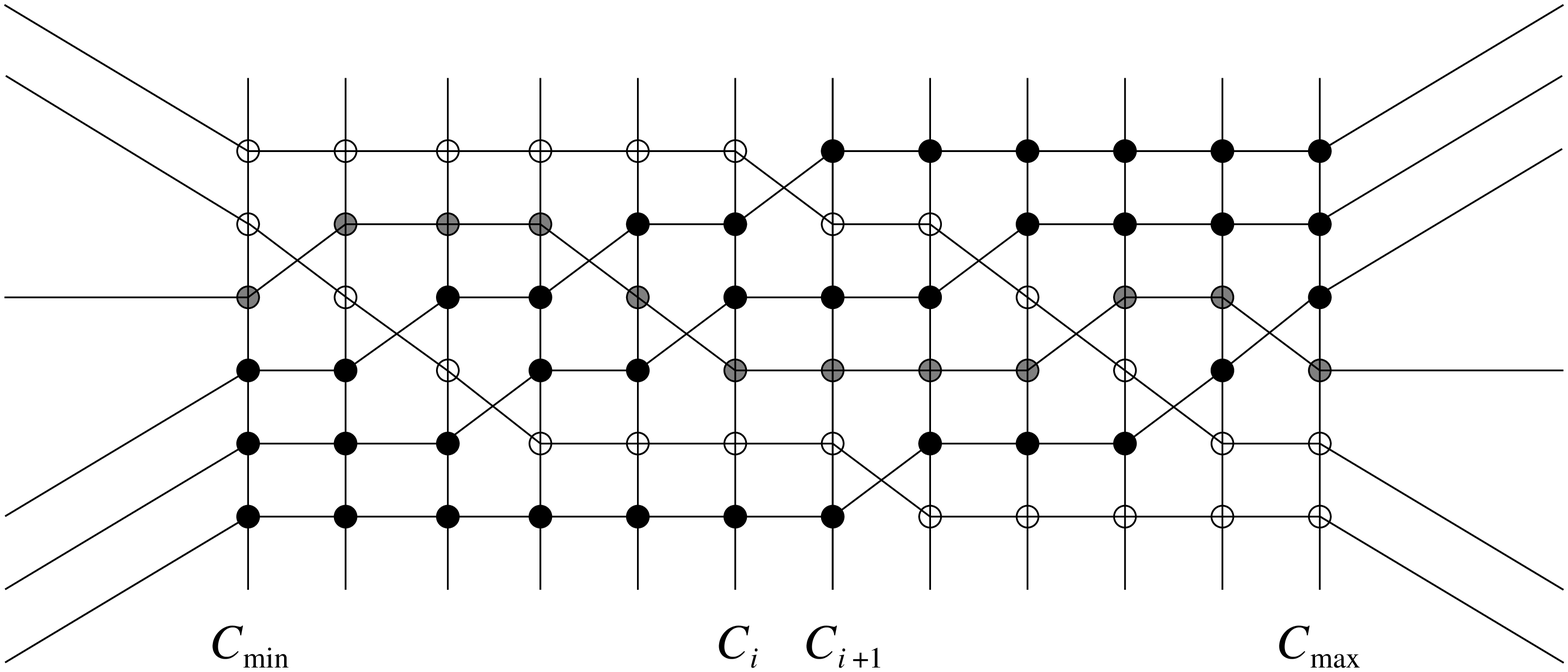,width=8cm}}~\hspace{10mm}~\parbox{3cm}{\psfig{figure=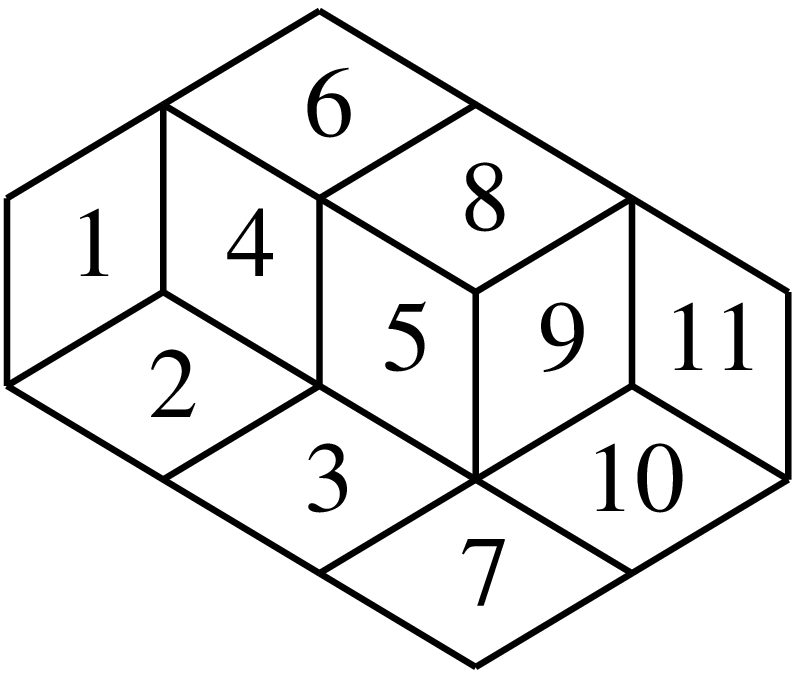,width=3cm}} \
\end{center}
\caption{Left: A $D-1 \ra 2$ grid and the successive sections $C_i$ by
a line of the $D$-th family. Each section can be seen as a $D-1 \ra 1$
tiling where the different species of tiles are represented by dots of
different colors. Two successive sections differ by a single flip,
that is the exchange of two tiles. In this example, $D=4$, $k_1=3$,
$k_2=1$ and $k_3=2$. Right: The corresponding labeled dual 
tiling.}
\label{ex31_32}
\end{figure}

Therefore the sequence $(P_i)$ is coded by a sequence $(C_i)$ of such
tilings, as illustrated in figure~\ref{ex31_32}. Since the same
argument can be applied to any subgrid $\GC$ of this $D \ra d$
problem, all the walks counted by the sum of the coefficients $a_j$ of
the counting polynomial can be seen as such sequences $(C_i)$.

It is now rather clear that all the configurations $C_0=O$ and
$C_K=S_0$ are the same for the partition problems on all subgrids
$\GC$: they are the two configurations where all the tiles of a same
family are adjacent and where the families are ordered according to
the grid configuration at infinity, which does not depend on $\GC$,
but only on the vectors $\vect{u}_i$. These two configurations will be
denoted by $\minindex{C}$ and $\maxindex{C}$, as in
figure~\ref{ex31_32}. Likewise, two successive tilings of a sequence,
$C_i$ and $C_{i+1}$, only differ by a tile flip, that is the exchange
of two adjacent tiles, since the two successive sections ``surround''
a single vertex of $\GC$. And there is a natural order between
two successive such tilings: considering how tiles are ordered on
$\minindex{C}$ and $\maxindex{C}$, it is clear which tiling should
come first in the sequence.

Conversely, let us establish that such a sequence $(C_i)$ of $D-1 \ra
1$ tilings, going from $\minindex{C}$ to $\maxindex{C}$, and where two
successive tilings only differ by a single flip and respect the above
order, contributes towards the sum of the coefficients $a_j$ of a
unique partition problem on a $D-1 \ra 2$ membrane~-- and therefore
towards the sum of the coefficients $a_j$ of the global $D \ra 2$
problem.  The proof is rather straightforward and is also illustrated
in figure~\ref{ex31_32}: considering two successive tilings, $C_i$ and
$C_{i+1}$, the two dots that represent the two flipping tiles of
different families are joined by two crossing segments; the
so-obtained vertex is labelled by $i+1$; all the other tiles are
joined by horizontal segments. Then one reconstructs a complete $D-1
\ra 2$ grid, the vertices of which are labeled by numbers increasing
on each de Bruijn line. This is precisely the kind of object counted
by coefficients $a_j$. 

Note that all the walks counted by $\sum a_j$
have the same length, since all the subgrids $\GC$ have the same
number of intersections.

\section{Number of maximal walks in a $n \ra 1$ configuration space}
\label{somme.des.aj}

In this section, we derive the number $A_n(k_1,\ldots,k_n)$ of walks
$(P_i)_{i=0,\ldots,K}$ in the $n \ra 1$ configuration space. Note that
for sake of simplicity, we note $n$ instead of $D-1$. We shall prove
that:
\begin{eqnarray}
\label{super-hips}
A_n(k_1,\ldots,k_n) & = & (\sum_{1\leq i<j \leq n} k_i k_j)! \\ 
 & &
 \times \prod_{1\leq i<j \leq n} { \fact2{(K_{ij}+k_i-1)}
 \fact2{(K_{ij}+k_j-1)} \over \fact2{(K_{ij}+k_i+k_j-1)}
 \fact2{(K_{ij}-1)}} \nonumber
\end{eqnarray}
where $K_{ij}=2(k_{i+1}+\ldots+k_{j-1})$ when $i < j$. 

We give two proofs: a purely algebraic one in the general case, using
results by Stanley~\cite{Stanley84}, and a combinatorial
one~\cite{These}. Note that the case where all the parameters $k_i$
are equal to 1 was already treated by Stanley~\cite{Stanley84}, and
that Edelman and Greene derived a nice combinatorial proof in this
case~\cite{Edelman87}.  We suppose that the notions of Young tableaux
and standard Young tableaux are known\footnote{A Young tableau is a
stacking of square boxes on a line, as displayed in figure~\ref{Young}
(left), with the only constraint that the number of boxes in columns
decreases from left to right. The shape of the tableau is the
decreasing sequence of the column heights.  A standard Young tableau
of a given shape is simply a numbering of the cells of the tableau by
integral numbers, running from 1 to the number of cells and increasing
in rows and columns. Figure~\ref{Young} provides examples. For a
presentation in relation with the symmetric groups $S_n$, the reader
can refer to \cite{Sagan}.}.

But before all, we need to introduce the relation between the tilings
considered in this paper and a class of computer science objects, the
so-called {\sl sorting algorithms}. This analogy will help the
presentation of the algebraic proof and will be useful in the
combinatorial proof.

\subsection{Tilings and primitive sorting algorithms}

In the sorting language, a {\sl comparator} $[i;j]$ acts on a list
$(x_1, x_2,\ldots, x_n)$ of numbers as follows: $x_i$ and $x_j$ are
respectively replaced by $\min(x_i,x_j)$ and $\max(x_i,x_j)$.
Following Knuth \cite{Knuth92}, we call a {\sl complete} sorting
algorithm a sequence of such comparators which sorts in the increasing
order any list of real numbers $(x_1,x_2,\ldots,x_n)$. This sorting
algorithm will be called {\sl primitive} if each comparator can be
written $[i,i+1]$. We also suppose that this algorithm is not
redundant, that is to say it does not contain any comparator $[i,j]$
that could be suppressed because previous comparators already insure
that $x_i \leq x_j$. Knuth shows that a sequence of comparators is a
sorting algorithm if it correctly sorts the completely reversed list
$(n,n-1,\ldots,1)$. This means that a complete primitive sorting
algorithm is a sequence of comparators $[i,i+1]$ that
transforms the list $(n,n-1,\ldots,1)$ into the list $(1,2,\ldots,n)$.

\begin{figure}[ht]
\begin{center}
\ \psfig{figure=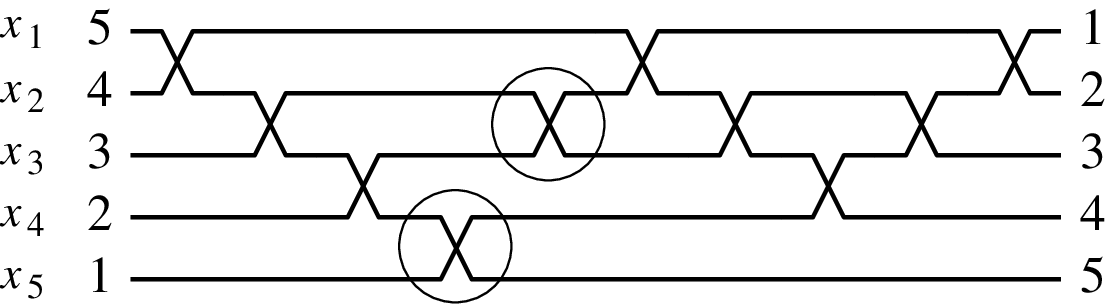,width=9cm} \
\end{center}
\caption{A diagram associated with a sorting algorithm acting on five
  element lists. A line follows a number during the sorting process.
  Each pair of lines cross, only once. The two circled comparators 
  can be exchanged without changing the corresponding tiling.}
\label{sort}
\end{figure}

Such an algorithm can have a diagrammatical representation as follows:
the $n$ variables $x_i$ are represented by $n$ horizontal lines. Each
comparator $[i,i+1]$ is represented by a crossing 
\begin{tabular}{c} \psfig{figure=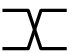} \end{tabular}
between lines $i$ and $i+1$. Figure~\ref{sort} illustrates this
construction. A continuous line follows a number during the sorting
process. For example, the greatest number is on the top at the
beginning and in the bottom at the end. Since every number must be
compared to every other one, and since there is no redundancy, there
are $\simp{n}{2}$ crossings.

We are now able to establish the link between those algorithms and $n
\ra 2$ tilings, more precisely with their de Bruijn representation.
Indeed, the analogy between the diagram of figure~\ref{sort} and a de
Bruijn grid with one line per family is straightforward: each
continuous line of the diagram represents a de Bruijn line and crosses
exactly once every other line.

However, there is a fundamental difference between both systems:
different sorting algorithms can represent the same de Bruijn grid
since only the crossing topology is meaningful. For example, in
figure~\ref{sort}, the fourth and the fifth comparator ({\sl i.e.}
[4,5] and [2,3]) are applied in this order (these comparator are
circled in the figure). If they were applied in the reverse ordre, the
algorithm would be different whereas the de Bruijn grid would be the same.

Therefore we are led to define equivalence classes of sorting
algorithms \cite{Knuth92,Elnitsky97}. We say that two comparators
$[i,i+1]$ and $[j,j+1]$ commute if $| i-j | >1$. Two algorithms are
equivalent if they differ by a finite number a comparator
commutations. These equivalence classes of $n$-element sorting
algorithms are in one-to-one correspondence with $n$-family grids with
one line per family, and therefore with tilings inscribed in polygons
of side 1. The number $A_n$ of equivalent classes has been computed by
Stanley~\cite{Stanley84} (see also Edelman and
Greene~\cite{Edelman87}) and is given by equation~\ref{super-hips} in
the case where $k_i=1$ for all $i$:
\begin{equation}
A_n(1,1,\ldots,1)={\simp{n}{2}! \over 1^{n-1} 3^{n-2} \ldots (2n-3)^1}.
\end{equation}

We need to generalize this point of view to systems with more than one
line per de Bruijn family, which leads to the definition of {\sl
partial} sorting algorithms. These algorithms are related to
pre-sorted lists of numbers. Indeed, let us suppose that we have $n$
families of $k_i$ numbers each ($i=1,\ldots,n$), and that in each
family, the numbers are already pre-sorted in the increasing
order. Then we are interested in the algorithms which order the whole
set of these numbers in the increasing order. We call them partial
sorting algorithms. The ideas are essentially the same as in the
previous case, except that, since the numbers of a given family are
already ordered, the corresponding lines do not need to cross. The
corresponding diagram is similar to a de Bruijn grid with $n$ families
of lines, $k_i$ lines in each family. The tilings are equivalence
classes of such algorithms.  They are inscribed in polygons of sides
$k_1,\ldots,k_n$.

\begin{figure}[ht]
\begin{center}
\ \psfig{figure=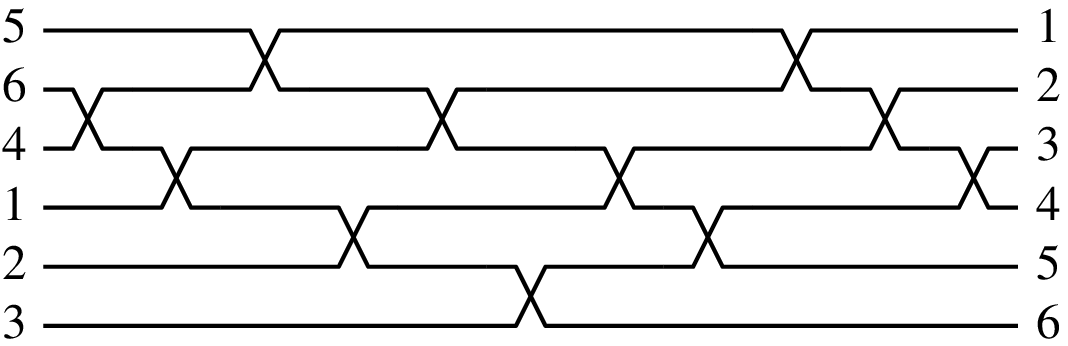,width=9cm} \
\end{center}
\caption{A partial sorting algorithm in the case of $n=3$ families,
containing respectively $k_1=3$, $k_2=1$ and $k_3=2$ pre-sorted
elements. The sequence $w_0$ appears vertically on the left of the
figure.}
\label{diagram.partial}
\end{figure}

In the partial sorting case, the reference list to be reversed
is not $(n,n-1,\ldots,1)$ any longer but a list $w_0$
were some elements are already sorted. If $\kappa_i=k_1+k_2+\ldots+k_i$,
then 
\begin{eqnarray}
w_0 & = & (\kappa_{n-1}+1,\ldots,\kappa_{n-1}+k_n, \\ \nonumber & & \
 \ \
 \kappa_{n-2}+1,\ldots,\kappa_{n-2}+k_{n-1},\ldots\ldots,1,2,\ldots,k_1).
\end{eqnarray}
There are $n$ pre-sorted blocks of $k_i$ elements each. An example
is provided in figure~\ref{diagram.partial}.

\medskip

Note also that a sorting algorithm can be seen as an ordering (or a
labeling) of the comparators of the corresponding equivalence class,
since the comparators come in a natural order: the labels are
increasing on each de Bruijn line, from left to right. As a
consequence, sorting algorithms are in one-to-one correspondence with
labeling of tilings, as defined in section~\ref{decomp.simp}
or~\ref{somme.des.aj.gen}.  This point is striking when comparing
figures~\ref{ex31_32} and \ref{diagram.partial}. Therefore the sums of
the coefficients $a_j$ of a $D \ra 2$ problem is equal to the number
of partial sorting algorithms of the suitable pre-sorted $D-1$
families of variables. This point will be helpful in the following
section.

\subsection{Walks and symmetric groups $S_n$ (algebraic proof)}

The main result used in this section is a theorem by Stanley (theorem
4.1 of reference~\cite{Stanley84}, and its corollary 4.2). We first
need to expose these results and to translate them in terms of our
notations and definitions. It is the object of the first
paragraph. Then, using the above equivalence between coefficients
$a_j$ and partial sorts, we shall apply Stanley's theorem to the case
of interest here, and derive relation~\ref{super-hips}.

\subsubsection{Stanley's theorem:}

The symmetric group $S_n$ of permutations on $n$ elements
is generated by the transpositions $\sigma_i=(i,i+1)$: any
permutation $w$ can be decomposed in products of transpositions.
A decomposition is said to be {\sl minimal} if,
using the relations between the generators $\sigma_i$:
\begin{equation}
\sigma_i^2=1 \ \ \mbox{ and } \ \ \sigma_i \sigma_{i+1} \sigma_i = 
\sigma_{i+1} \sigma_i \sigma_{i+1},
\label{relations}
\end{equation}
it cannot be simplified into a shortest decomposition; then all the
reduced decomposition of $w$ have the same length\footnote{The length
of a decomposition is the number of generators necessary to write
it.}, denoted by $l(w)$. If $w=\left( \begin{array}{cccc} 1 & 2 &
\cdots & n \\ a_1 & a_2 & \cdots & a_n \end{array} \right)$, we define
\begin{eqnarray}
r_i(w)& = & \card\{j:j<i \mbox{ and } a_j > a_i \} \\
s_i(w)& = & \card\{j:j>i \mbox{ and } a_j < a_i \}.
\end{eqnarray}
Then $\lambda(w)$ is the sequence obtained by arranging the
numbers $r_i(w)$ in descending order (and ignoring any 0's); $\mu(w)$
is the {\sl conjugate}\footnote{If a decreasing sequence is drawn like
a Young tableau, its conjugate is the decreasing sequence of row
lengths. For example, the conjugate to (3,2,2,1) is (4,3,1).} to the
sequence obtained by arranging the numbers $s_i(w)$ in descending
order.

Stanley~\cite{Stanley84} states that: {\em if $\lambda(w)=\mu(w)$,
then the number of reduced decompositions of $w$ is equal to the
number $f^{\lambda(w)}$ of standard Young tableaux of shape
$\lambda(w)$.}

\subsubsection{Application to walks in $n \ra 1$ configuration spaces:}
We still denote by $w_0$ the permutation as defined in the previous
section. The idea is to see each comparator $[i,i+1]$ as a generator
$\sigma_i=(i,i+1)$ of a reduced decomposition and to identify a
sorting algorithm with a reduced decomposition of $w_0$.  More
precisely, the set of comparators can be seen as the generators of a
group, and they obey the same relations as the $\sigma_i$: they
satisfy relations~\ref{relations}, and $[i,i+1]$ and $[j,j+1]$ commute
if $| i-j | >1$. Finally, the non-redundant character of sorts is
equivalent to the reduced character of decompositions. As a
consequence, the number $A_n(k_1,\ldots,k_n)$ of partial sorts is
equal to the number of reduced decompositions of $w_0$.

\begin{figure}[ht]
\begin{center}
\ \psfig{figure=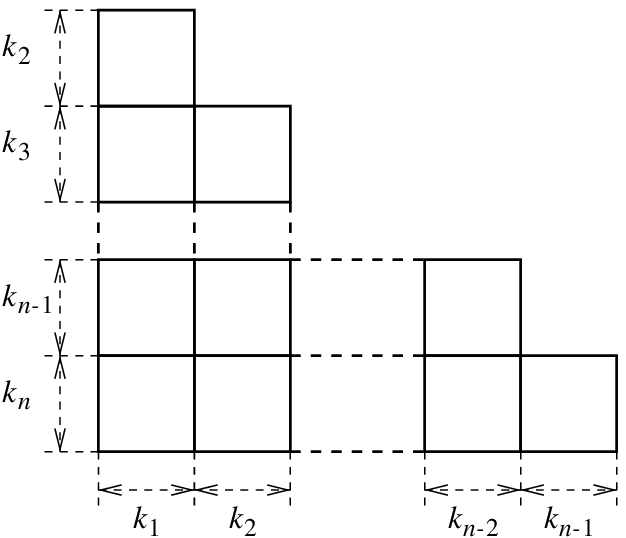,width=6cm} \
\end{center}
\caption{A Young tableau of shape $\lambda(w_0)=
\mu(w_0)$. The numbers $k_i$ denote the number of rows and columns in
each rectangular block. Such a tableau is called a block tableau
of size $(k_1,k_2,\ldots,k_n)$.}
\label{blocs}
\end{figure}

\begin{figure}[ht]
\begin{center}
\ \psfig{figure=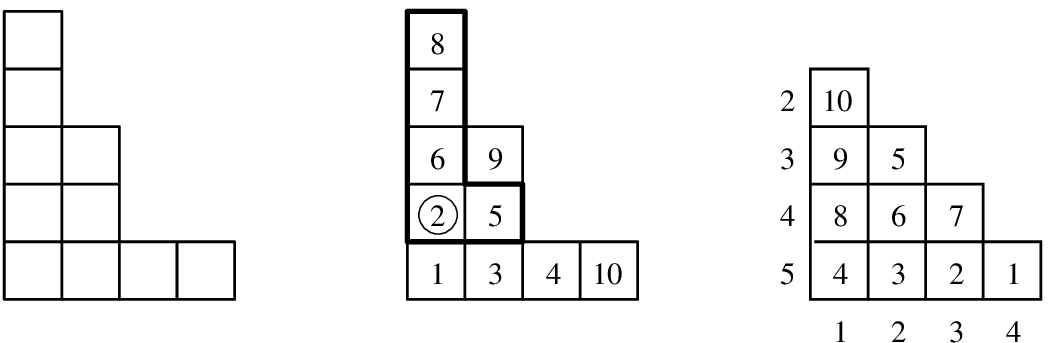} \
\end{center}
\caption{Left: a four-column Young tableau of shape
$\lambda=(5,3,1,1)$. Middle: an associated standard tableau. The hook
corresponding to the circled entry is represented. Its length is $h_c=5$.
The hook rank of this cell is $r_c=1$ and its hook height is 
$\theta_c=4$: this tableau is {\sl not} balanced.
Right: a balanced stair tableau of order $5$.}
\label{Young}
\end{figure}

Now, in order to apply Stanley's statement, we must compute the
quantities $\lambda(w_0)$ and $\mu(w_0)$, as defined in
reference~\cite{Stanley84}: in the present case, one gets
\begin{equation}
r_i(w_0) = k_n + \ldots + k_j \ \ \mbox { if } \ k_n + \ldots + k_j +1 \geq i 
	\geq k_n + \ldots + k_{j-1}
\end{equation}
\begin{equation}
s_i(w_0) = \kappa_{j-1} \ \ \mbox { if } \ k_n + \ldots + k_j +1 \geq i 
	\geq k_n + \ldots + k_{j-1}.
\end{equation}
Thus
\begin{equation}
\lambda(w_0)=\mu(w_0)=(\underbrace{k_n+\ldots+k_2}_{k_1
\ \mbox{\scriptsize times}}, 
\underbrace{k_n+\ldots+k_3}_{k_2
\ \mbox{\scriptsize times}} , \ldots ,
\underbrace{k_n}_{k_{n-1} \ \mbox{\scriptsize times}}),
\end{equation}
and Stanley's theorem applies. The Young tableau of shape
$\lambda(w_0)$ is represented in figure~\ref{blocs}. Such a tableau
will be called a {\sl block tableau} of size $(k_1,k_2,\ldots,k_n)$ in
the following.
The number $A_n(k_1,k_2,\ldots,k_n)$ of reduced decompositions of
$w_0$ is equal to the number $f^{\lambda(w_0)}$ of standard Young
tableaux of shape $\lambda(w_0)$. This number can be derived from
Young's {\sl hook-length formula} \cite{Young27,Greene79}: given a
shape $\lambda$, the {\sl hook} associated with a given cell $c$ of
the tableau is the set of cells above and at the right of $c$,
including $c$ itself (figure~\ref{Young}). It is denoted by $H_c$. The
{\sl hook length} $h_c$ is the number of cells in $H_c$. Then the
number $f^\lambda$ of standard Young tableaux of shape $\lambda$ is:
\begin{equation}
f^\lambda = {N! \over \prod_c h_c},
\end{equation}
where $N$ is the total number of cells and the product runs over all
cells.

This hook-length formula can know be applied to the above tableau: the
hook-length of the cell in the upper right corner of block $k_i \times
k_j$ is $K_{ij}+1$, from which one deduces all the hook-lengths on the
block and their product: $\displaystyle{{\fact2{(K_{ij}+k_i+k_j-1)}
\fact2{(K_{ij}-1)}} \over \fact2{(K_{ij}+k_i-1)}
\fact2{(K_{ij}+k_j-1)} }$. Moreover, $N=\sum_{i<j} k_i k_j$, from which
equa\-tion~\ref{super-hips} follows.

\subsection{Combinatorial proof}

In this section, we provide a combinatorial bijective proof of the
previous result~\cite{These}. This proof follows the same scheme as
the proof by Edelman and Greene~\cite{Edelman87} in the case where all
the parameters $k_i$ are equal to 1. We need first to introduce the
notion of {\sl balanced tableaux}~\cite{Edelman87}.

We consider a Young tableau, together with a labeling of its cells,
running from 1 to the number of cells, $K$. Note that now this tableau
is not necessarily standard, that is to say the labels are not
necessarily ordered in each row and in each column. Given a cell $c$
and its hook $H_c$, we define the {\sl hook rank} $r_c$ of $c$ as the
number of cells of $H_c$ whose labels are smaller or equal to the
label of $c$. We also define the {\sl hook height} $\theta_i$ as the
number of cells above $c$ (including $c$ itself). The tableau is said
to be balanced if for all cells $c$, $r_c = \theta_c$ (see
figure~\ref{Young}).

At last, a tableau of shape $(n-1,n-2,\ldots,1)$
(figure~\ref{Young}, right) will be called a {\sl stair tableau} (of
order $n$). Note that a stair tableau is a particular case of block
tableau, as defined in the previous section.

In the case $k_i=1$, the situation is as follows: Edelman and Greene
build a bijection between {\sl complete} sorting algorithms on $n$
elements and balanced block tableaux of order $n$ and then a bijection
between those balanced tableaux and standard tableaux of order $n$,
which are then enumerated {\em via} the hook-length formula. In the
following, we use and generalize these results to the case of partial
sorts. In fact, in reference~\cite{Edelman87}, the authors also
establish the bijection between balanced tableaux of any shape and
standard tableaux of the same shape. As a consequence, we only need to
generalize the first bijection between {\sl partial} sorting
algorithms on $n$ families containing $k_i$ elements each and balanced
block tableaux of size $(k_1,\ldots,k_n)$, as in
figure~\ref{blocs}. Once this correspondence is established, the rest
of the proof is based upon the hook-length formula, as in the previous
algebraic proof. The rest of the section is devoted to this 
correspondence.

To begin with, let us consider a {\sl complete} sorting algorithm on
$n$ lines: it is a sequence of crossings between de Bruijn
lines. These crossings are labeled by integers running from 1 to
$K=\simp{n}{2}$, from left to right. Following Edelman and Greene, the
intersection label of two lines indexed by $a$ and $b$ is denoted by
$t_{ab}$.  In the stair tableau, this number is written in the cell
situated on the $a$-th column and on the $(n-b+1)$-th line (starting
from the bottom): in figure~\ref{Young} (right), the so-obtained
tableau corresponding to the complete sorting algorithm of
figure~\ref{sort} is represented.  Then it can be
proven~\cite{Edelman87} that this tableau is balanced and more
precisely that this construction establishes a bijection between both
classes of objects.

Let us now focus on partial sorting algorithms.  Since there are
couples of de Bruijn lines which do not intersect, it is rather
natural to consider block tableaux, which are stair tableaux where
some cells are missing. More precisely, we will consider partial
sorting algorithms and block tableaux as {\sl amputated} complete
sorting algorithms and {\sl amputated} stair tableaux,
respectively. The idea is to define {\sl canonical} amputations in
order to preserve Edelman and Greene's bijection between amputated
objects, as discussed below.

\begin{figure}[ht]
\begin{center}
\ \psfig{figure=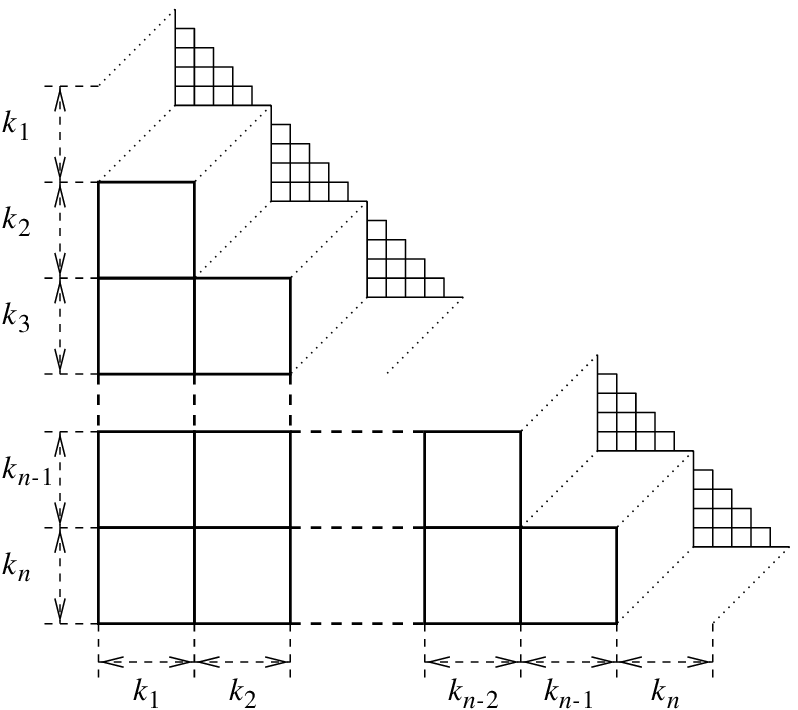,width=6.5cm} \
\end{center}
\caption{Amputation of a stair tableau.}
\label{amput.tab}
\end{figure}

Figure~\ref{amput.tab} illustrates the tableau amputation process: given $n$ 
integers $k_1,\ldots,k_n$ and a stair tableau of order $N=\sum k_i$, 
$n$ small stair tableaux of order $k_i$ are removed from the large
one in order to get a block tableau of size $(k_1,\ldots,k_n)$.

\begin{figure}[ht]
\begin{center}
\ \epsffile{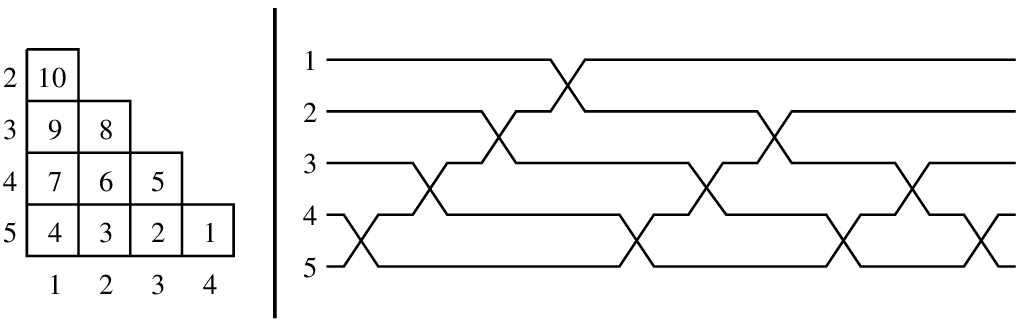} \
\end{center}
\caption{A canonical complete sort and the corresponding canonical
stair tableau.}
\label{canon}
\end{figure}

As far as the amputation of sorting algorithms is concerned, we first
need to define {\sl canonical} complete sorts: the simplest way to
characterize them is by their corresponding stair tableaux, the cells
of which are increasing from left to right and from bottom to top, as
illustrated in figure~\ref{canon}. They are usually referred as
``bubble-sorts'' in the literature.

\begin{figure}[ht]
\begin{center}
\ \psfig{figure=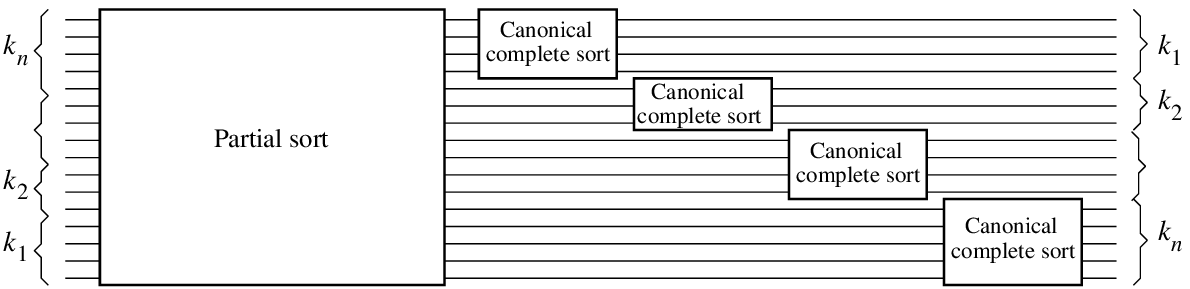} \
\end{center}
\caption{A complete sequential sorting algorithm: it is a partial
sorting algorithm followed by $n$ canonical complete sorts on $k_i$
elements.}
\label{amput.tri}
\end{figure}

If we consider now a partial sorting algorithm with $k_i$ elements in
each family, it can be canonically transformed into a complete sort:
we simply add $n$ canonical complete sorts on $k_i$ elements at its
end, as in figure~\ref{amput.tri}. These sorts appear in the order of
their indices $i$. The so-obtained sort is very particular since it
ends with $n$ canonical sorts. Therefore it will be called a {\sl
sequential} complete sort of order $(k_1,\ldots,k_n)$.  By
construction, there is a one-to-one correspondence between sequential
complete sorts of order $(k_1,\ldots,k_n)$ and partial sorts of the
same size.

Let us now characterize the structure of the Young tableau $\tau$
associated with such a sequential sort. The
$\simp{k_1}{2}+\simp{k_2}{2}+\ldots+\simp{k_n}{2}$ last intersections
in the sort are those appearing in the $n$ canonical complete
sorts. It is easily checked that the corresponding labels in the large
stair tableau appear in the $n$ small canonical stair tableaux
involved in the amputation process. As a consequence, the labels of
the amputated tableau run from 1 to $K$, where $K$ is its number of
cells. These labels code the $K$ intersections of the partial sort
remaining of the original sequential complete sorting algorithm.

\begin{figure}[ht]
\begin{center}
\ \psfig{figure=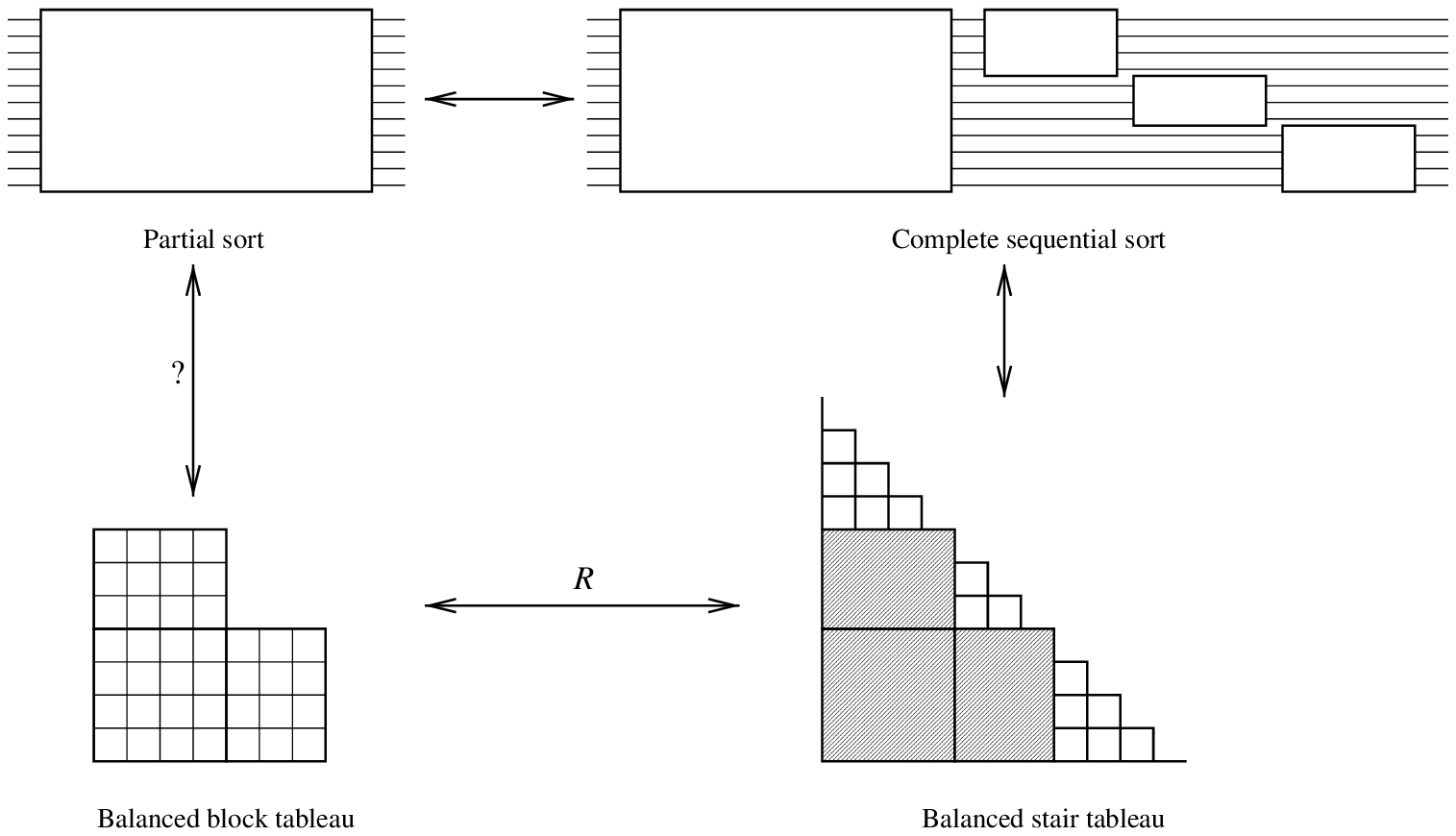,width=12cm} \
\end{center}
\caption{Sequence of bijections establishing the one-to-one
correspondence between partial sorting algorithms on $n$ families with
$k_i$ numbers each and balanced block tableaux of size
$(k_1,\ldots,k_n)$. In the lower right stair tableau, the grayed
sub-tableau is a pre-balanced block one. The bijection $R$ puts such
tableaux in one-to-one correspondence with balanced block tableaux of
the same shape.}
\label{bij}
\end{figure}

To sum up, as it is illustrated in figure~\ref{bij}, starting from
partial sorts, we biunivocally construct complete sequential sorts,
then balanced tableaux, the $K$ last labels of which are situated in
the $n$ stair sub-tableaux. When these sub-tableaux are removed from
the large one, we obtain a class of block tableaux of size
$(k_1,\ldots,k_n)$, which will be called {\sl pre-balanced} tableaux.
Remember now that are goal is to establish a bijection between partial
sorts and balanced block tableaux. Thus we need to construct 
a bijection (denoted by $R$ in figure~\ref{bij}) between those
pre-balanced tableaux of size $(k_1,\ldots,k_n)$ and balanced block
tableaux of the same size.

\medskip

\noindent {\bf Bijection $R$:} actually, $R$ is an involution: in each
group of $k_i$ columns, it inverses the order of columns; for example,
the first column becomes the $k_1$-th one, the second one becomes the
$(k_1-1)$-th one and so on. Likewise, the $(k_1+1)$-th column becomes the
$(k_2-1)$-th one. There is a rather deep reason for such a column
permutation: the $n$ canonical complete sorts added at the end
of a partial sort in order to make it complete also reverse the
order of lines in each group of $k_i$ lines. The role of $R$ is 
to keep track of this fact. It is now a rather technical task to
prove that $R$ provides balanced tableaux and that it is a bijective
map.

\medskip

{\small We shall temporarily admit the following results which will
be proven at the end of this section: in a pre-balanced tableau, in
each block $k_i \times k_j$, the labels are decreasing in each 
line and column (from bottom to top); in a balanced block tableau,
they are increasing in lines and decreasing in columns in such
a block. 

Consider a pre-balanced block tableau and in this tableau a hook $H_c$
associated with the cell $c$ situated in the block $k_i \times k_j$,
and in this block in the line $u$ (from bottom to top) and column $v$
($1 \leq u \leq k_i$ and $1 \leq v \leq k_j$). This hook comes
from a larger one, $H'_c$, in the stair tableau which has been
amputated. In the process the hook has lost $C$ cells, $C_1$ on its
right and $C_2$ on its top (see figure~\ref{HH'}). Thus the height of
$H_c$ is equal to 
$$\theta_c=\theta'_c-C_2.$$
On the other hand, all the lost cells, coming from the $n$ removed
stair tableaux, had labels larger than the label of the cell $c$. As a
consequence, the hook rank remains unchanged:
$$
r_c=r'_c.
$$

\begin{figure}[ht]
\begin{center}
\ \psfig{figure=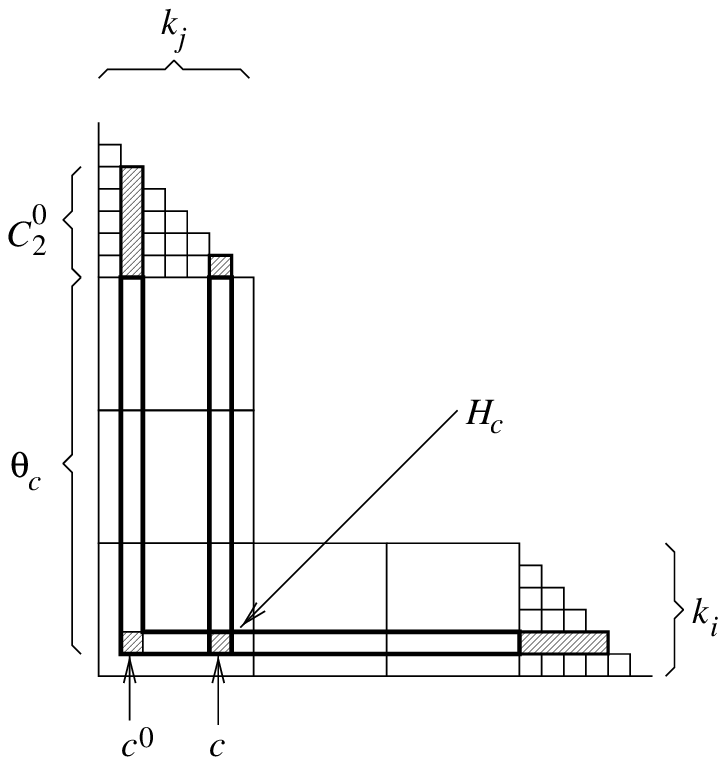,width=6cm} \
\end{center}
\caption{A hook $H_c$ of a pre-balanced tableau. As compared to the
original hook $H'_c$ (before the amputation), it has lost cells on its
top and on its right (dashed). By the involution $R$, $c$ receives the
label of the cell $c^0$.}
\label{HH'}
\end{figure}

Now, $R$ permutes the labels of the block tableau but does not change its
shape. In the process, the cell $c$ receives the label of the
cell $c^0$ still situated in the line $u$, but in the column
$k_j-v+1$.

For a given quantity $A$ assigned to each cell, we denote by $A^0$ the
value of this quantity for the cell $c^0$ in the pre-balanced tableau,
and $A^R$ its value in its image by $R$. In particular,
$\theta_c^R=\theta_c^0$, and we have just proven that
$r_c^0-\theta_c^0=C_2^0$. Moreover, $r^R_c=r^0_c-(k_j-v^0)$. Indeed,
by $R$, the hook $H_c$ receives the labels of the hook $H_{c_0}$
outside the block $k_i \times k_j$; and in the block, the labels,
which were decreasing in the line $u$ of the pre-balanced tableau, are
increasing in its image by $R$. As a consequence,
$$
r^R_c - \theta^R_c = C_2^0 - (k_j - v^0).
$$
Now $C_2^0 = (k_j - v^0)$ by definition. Therefore $r^R_c =
\theta^R_c$ and the tableau is balanced. Conversely, given a balanced
tableau, since the labels are increasing in each line of each block,
one proves that its image by the involution $R$ is pre-balanced. We
have established the bijection.

We need to prove the two above assertions about the order of labels in
blocks of balanced and pre-balanced tableaux. We only give sketches of
the proofs. As far as pre-balanced tableaux are concerned, the proof
is rather straightforward: a block $k_i \times k_j$ contains labels
associated with all the intersections of two families of lines. If
those lines are isolated from the rest of the tiling, it becomes clear
that the order in which intersections occur is constrained. For
balanced tableau, the proof is more complex. The basic idea is to
construct a proof by ``planar induction'': we prove that if a suitable
$\PC$ property is true for cells above and at the right of a given
cell $c$, then it is also true for $c$. Then if $\PC$ is true for
cells on the upper right corners of the tableau, it will be true for
every cell. In the present case, if $t_c$ still denotes the label of
the cell $c$, then the property reads:

\medskip

\noindent {\em $\PC(c)$: in the block $k_i \times k_j$ to which $c$
belongs, the cells above $c$ have labels smaller than $t_c$, while the
cells at its right have labels greater than $t_c$.}

\medskip

\noindent A proper use of the balanced character of a block tableau
proves that it satisfies the above planar induction principle.}

\section{Proofs of section \protect{\ref{results}}}
\label{proofs4}
In this appendix, we only give sketches of proofs for the results
used in section~\ref{results}. The complete proofs can be found in the
relevant references.

\subsection{Elnitsky's formulas}
\label{Elnitsky}

\begin{figure}[ht]
\begin{center}
\ \psfig{figure=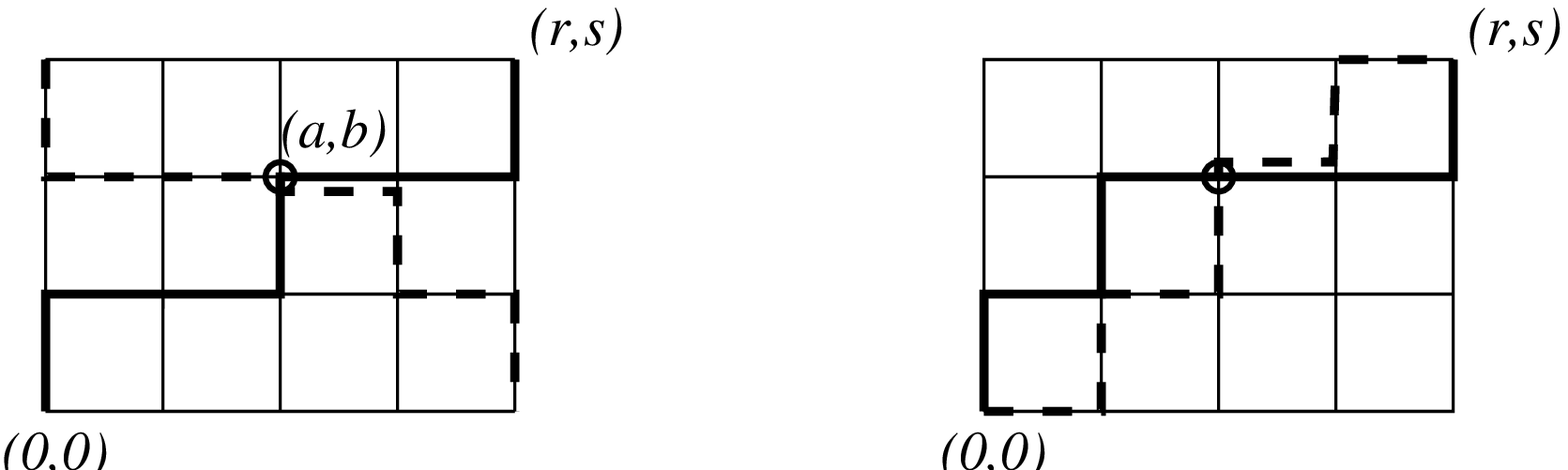,width=9cm} \
\end{center}
\caption{Two lattice-path diagrams, associated with the two tilings of
figure~\protect{\ref{ex42}}. The de Bruijn lines of the two
single-line families are represented by a pair of paths running on a
square lattice.  These paths go from one corner to the diagonally
opposite one. In case of ambiguity, the vertex where the de Bruijn
lines intersect is distinguished by a circle. There are two cases:
either the two sides of length one of the polygonal boundary are
adjacent (right) or not (left).}
\label{sq.grids}
\end{figure}

In order to prove relations~\ref{Elnit1} and \ref{A7}, we need, in the
octagonal case, a slightly different representation of de Bruijn grids
than those presented in the introductory section: as displayed in
figure~\ref{sq.grids}, the lines of the two first families form a
square lattice (of sides $r$ and $s$ in this case), on which the lines
of the third and fourth de Bruijn families run: they are represented
by directed walks on the lattice, going from one corner to the
diagonally opposite one. According to whether the two sides of length
one are adjacent or not, the two paths have the same starting and
ending points (right) or not (left).  In this representation, the de
Bruijn line intersections must be distinguished to avoid possible
ambiguities due to path tangency (it will be called the {\em
distinguished} vertex in the following).  For example, the octagonal
tilings of figure~\ref{ex42} are represented by the pair of paths of
figure~\ref{sq.grids}.

The proof of relation~\ref{Elnit1} is now straightforward: if $(a,b)$
are the coordinates of the distinguished vertex in the grid, then the
number of configurations is the product of the 4 number of choices for
the four pieces of paths going from $(a,b)$ to the four corners, that is
the product of four binomial coefficients. Now summing over all
$(a,b)$ configurations one gets relation~\ref{Elnit1}.

\medskip

The proof of relation~\ref{A7} is a little more complex and involves
some modifications of the pair of paths, as usual in this kind of
calculation. The idea is to exchange the two paths {\sl after} the
distinguished vertex, in order to get non-crossing (but possibly
touching) paths, and then to shift some parts of those paths in order
to get a non-touching pair. The latter pairs can be counted with help
of the determinental Gessel-Viennot method~\cite{Gessel85}. The
interested reader will refer to reference~\cite{Elnitsky97} for more
details.

\subsection{Brock's recursion relation}
\label{Brock}

\begin{figure}[ht]
\begin{center}
\ \psfig{figure=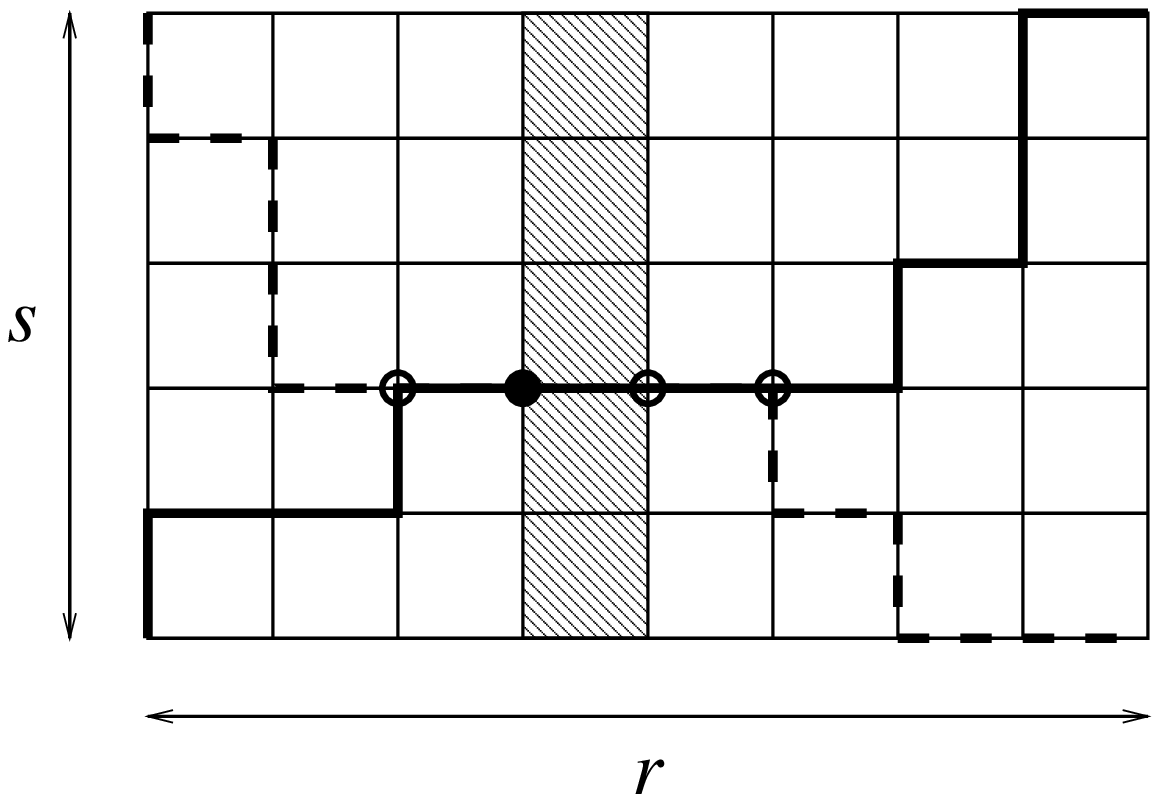,width=5.5cm} \
\end{center}
\caption{When the two path intersection is a segment (vertical or
horizontal, of length $l_p > 0$; here $l_p=3$), the same pair can
represent several tilings, depending on the position of the
distinguished vertex. If this point is not the last possible one ({\em
i.e.} the point with the highest coordinates), then a strip of the
lattice can be removed (dashed in this figure, at the right of
the black distinguished vertex). The so-obtained
diagram is associated with a smaller tiling ($r-1 \times s$ in
the present case; $r \times s-1$ if the segment were vertical).}
\label{Brock.proof}
\end{figure}

Let us now focus on the recursion relation~\ref{brock}. We use again
the lattice-path representation, as defined in the previous section.
Given a pair $p$ of crossing paths, with no {\em a priori} distinguished
vertex, let us denote by $l_p$ the length of their intersection
(see figure~\ref{Brock.proof}). Note that
this intersection can be a point ($l_p=0$), a horizontal or a vertical 
segment (of length $l_p$). If $\PC_{r,s}$ is the set of pairs,
then 
\begin{equation}
W^{4 \ra 2}_{r,1,s,1} = \sum_{p \in \PC_{r,s}} l_p+1 = \Simp{r+s}{r}^2
+ \sum_{p \in \PC_{r,s}} l_p,
\end{equation}
since there are $l_p+1$ possible choices for the distinguished vertex and
there are $\Simp{r+s}{r}^2$ such pairs of paths.

\medskip

In the case where $l_p > 0$, if the distinguished vertex is not the
point with the highest coordinates ($l_p$ possibilities), as in
figure~\ref{Brock.proof}, then a vertical or horizontal strip of the
square lattice can be removed without changing the nature of the
diagram: it is the strip of width 1, at the right of (respectively
above) the distinguished vertex if the intersection is horizontal
(respectively vertical).  As a result, one still have a square lattice
(but one of its side lengths is lowered by 1) with a pair of paths
(but the length of their intersection has been lowered by one in the
process). That is why
\begin{equation}
\sum_{p \in \PC_{r,s}} l_p = \sum_{p \in \PC_{r-1,s}} (l_p+1) +
\sum_{p \in \PC_{r,s-1}} (l_p+1).
\end{equation}
In conclusion,
\begin{eqnarray}
W^{4 \ra 2}_{r,1,s,1} & = & \Simp{r+s}{r}^2
+ \sum_{p \in \PC_{r,s}} l_p \\ \nonumber
 & = & \Simp{r+s}{r}^2
+ \sum_{p \in \PC_{r-1,s}} (l_p+1) +
\sum_{p \in \PC_{r,s-1}} (l_p+1) \\ \nonumber
 & = & \Simp{r+s}{r}^2 + W^{4 \ra 2}_{r-1,1,s,1} +
W^{4 \ra 2}_{r,1,s-1,1}, \\ \nonumber
\end{eqnarray}
which achieves the proof.

\section{What about the general $D \ra d$ case?}
\label{cycles}

In this last appendix, we discuss what we know and what we do not know about
the general $D \ra d$ case which was more widely studied in
reference~\cite{These}. The different points and results tackled
in the present paper are discussed. It will become clear that
the possible existence of cycles in the partition-on-tiling problems
is the major obstacle to simple generalizations.

\subsection{Partition-on-tiling point of view and configuration space}

As far as partitions on tilings are concerned, all that as been said
in the octagonal case can be transposed to the general case: a fixed
boundary $D \ra d$ tiling~\cite{Bibi97,These} can be coded in a single
way as a partition on a $D-1 \ra d$ tiling. The de Bruijn lines are
still defined as lines joining together the middles of opposite faces
of rhombic tiles, and the parts of those partitions are still
increasing along such lines. The reader can refer to
Bailey~\cite{Bailey97} for a more formal treatment of
this question. Note that in this case, there also exist de Bruijn
families of {\sl hyper-surfaces}, associated with an edge orientation.

However, as it was suggested in reference~\cite{Bibi97}, the geometry
of configuration spaces might be more complex beyond the octagonal
case. Indeed, among the order relations $x_i \geq x_j$ between the parts of
a partition-on-tiling problem, nothing forbids {\em a priori} the
existence of cycles of inequalities, such as $x_{i_1} \geq x_{i_2}
\geq \ldots \geq x_{i_q} \geq x_{i_1}$, which enforces all these
variables to be equal.

At least two examples of $6 \ra 3$ tilings (in relation with $7 \ra 3$
tiling problems) are known which display such cycles. The first one
can be found in reference \cite{Bjorner93} (example 10.4.1) and the
second one in \cite{Sturmfels93} (example 3.5)\footnote{Note that in
these references (proposition 10.5.7 of \cite{Bjorner93} and corollary
4.5 of \cite{Sturmfels93}), it is also stated that no such cycles
exist in two-dimensional tilings.}. In these examples, the tilings are
defined by their dual de Bruijn grids~-- which are families of
2-dimensional de Bruijn surfaces in a 3-dimensional space~-- together with an
orientation of de Bruijn lines.

When such a cycle exists, all the parts of the cycle have a
``collective behavior'', which is not compatible with the previous
description: they behave like a single {\sl effective} part. In
particular, the number of effective parts in this partition problem,
denoted as $K'$, is strictly smaller than $K$.  Thus the counting
polynomial of this partition problem becomes:
\begin{equation}
\sum_{j=0}^{M'} a_j^{K'} \Simp{K'+p-j}{K'},
\label{K'}
\end{equation}
The counting polynomial of the whole tiling problem is a sum of such
polynomials, with possibly many different $K'$.

Moreover, the existence of cycles invalidates the proof of the
connectivity of the configuration space (section~\ref{config.space}).
As far as we know, this point is an open question in the general $D
\ra d$ case. Note however that whenever one can prove that order
relations on fibers contain no cycles, then the configuration space is
connected.

\subsection{Decomposition in simplices~-- descent theorem}

We prove that the existence of cycles does not alter the previous
results about the sums of coefficients $a_j$ and walks in
configuration spaces, provided these objects are suitably defined: if
we focus only on coefficients $a_j^K$ associated with configuration
spaces of effective dimension $K$ (and not $K'<K$), then the sum
$\Sigma^K$ of these coefficients is equal to the number of maximal
walks in the corresponding $D-1 \ra d-1$ configuration space.  Note
that this quantity $\Sigma^K$ still characterizes the leading
coefficient of the counting polynomial (of degree $K$) as $p$ goes to
infinity.

More precisely, if there exists a cycle in the partition problem on a
membrane, we have just seen that the coefficients $a_j^{K'}$ of this
problem do not contribute to $\Sigma^K$. On the other hand, let us
consider a step in a walk in the $D-1 \ra d-1$ configuration space: we
have seen that the part $x$ which differentiates the two consecutive
configurations of this step is such that all parts lesser (resp.
greater) than $x$ in the graph are equal to 1 (resp. 0). This part is
equal to 0 in a partition and to 1 in the other one. As far as the
parts of the cycle are concerned, since they are all equal, they
cannot but jump from 0 to 1 all together, which is not conform to our
definition of walk in the configuration space, in terms of {\sl single}
elementary flips. Conversely and for similar reasons, a maximal walk
in the configuration space cannot give a membrane with cycles.

In conclusion, as well $\Sigma^K$ as the number of walks are not
concerned by partitions with cycles, and the above result remains
valid.

\begin{figure}[ht]
\begin{center}
\ \psfig{figure=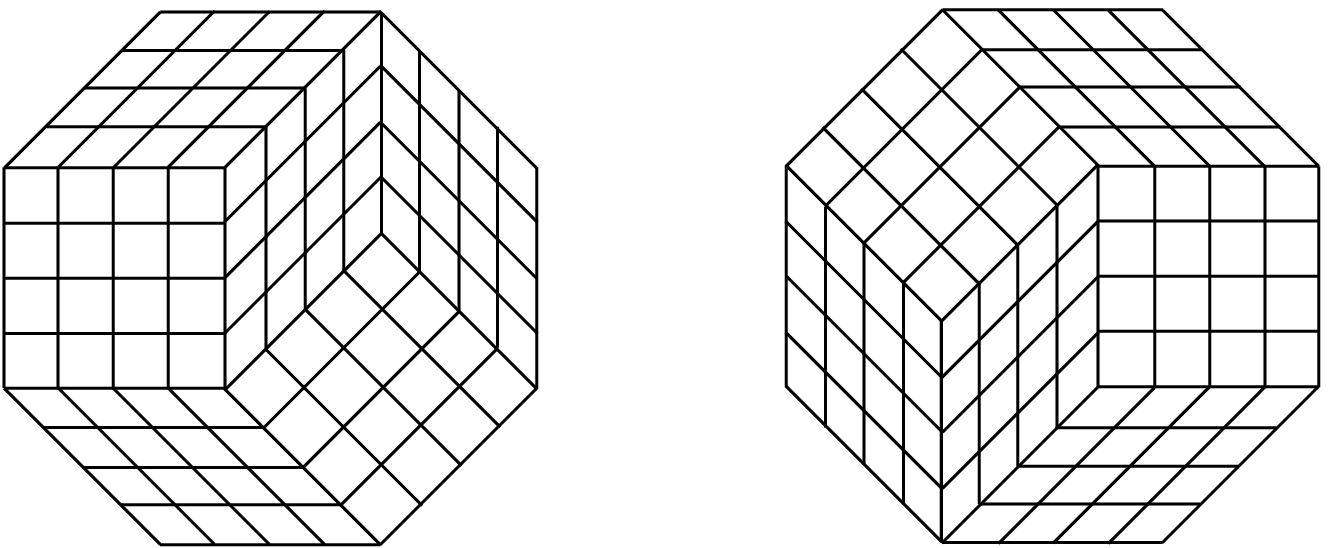,height=3cm} \
\end{center}
\caption{\label{macle} Two examples of extremal $4 \ra 2$ tilings
displaying a macroscopic faceting. Such tilings are extremal
configurations for walks associated with a $5 \ra 3$ problem.
There are four similar pairs of faceted tilings which could 
be chosen as extremal configurations, depending on how the
fourth de Bruijn family is chosen among the four possible ones.}
\end{figure}

Before going on, let us specify what the extremal tilings become in
larger dimension: $\minindex{C}$ and $\maxindex{C}$ are defined by
partitions where all the parts are equal to 0 and 1, respectively.
Therefore the corresponding tilings present a faceted aspect, as on
figure~\ref{macle}. Note that among the different possible faceted
tilings, the two extremal ones depend on how the $D$-th de Bruijn
family of surfaces is chosen.

At last, we recall that there cannot exist a descent theorem as simple
as the octagonal one in the general case. Indeed, its derivation is
closely related to the existence of a zero-descent simplex in each
partition-on-tiling problem, that is to the $K$-dimensional character
of the associated configuration space. But we have shown that this
point is not granted in general, since there can exist
partition-on-tiling problems for which the configuration space has a
dimension $K'$ smaller than $K$.


\section*{References}


\begin{thebibliography}{99}


\bibitem{Shechtman84} D. Shechtman, I. Blech, D. Gratias, J.W. Cahn,
Metallic phase with long-range orientational order and no
translational symmetry, {\em Phys. Rev. Lett.} {\bf 53}, 1951 (1984).

\bibitem{Penrose74} R. Penrose, The role of aesthetics in pure and
applied mathematical research, {\em Bull. Inst. Math. Appl.} {\bf 10},
226 (1974).

\bibitem{Bibi97} N. Destainville, R. Mosseri, F. Bailly,
Configurational entropy of codimension-one tilings and directed membranes, 
{\em J. Stat. Phys.} {\bf 87}, Nos 3/4, 697 (1997).

\bibitem{Kuo}  N. Wang, H. Chen, and K.H. Kuo, 
Two-dimensional quasicrystal with eightfold rotational symmetry,
{\em Phys. Rev. Lett.} {\bf 59}, 1010 (1987).

\bibitem{Li92} W. Li, H. Park, M. Widom, Phase diagram of a random 
tiling quasicrystal, {\em J. Stat. Phys.} {\bf 66},
Nos 1/2, 1 (1992).

\bibitem{Widom93} M. Widom, Bethe Ansatz Solution of the Square-Triangle 
Random Tiling Model, {\em Phys. Rev. Lett.} {\bf 70}, 2094 (1993).

\bibitem{Kalugin94} P.A. Kalugin, The Square-Triangle Random-Tiling
  Model in the Thermodynamic Limit, {\em J. Phys. A: Math. Gen.} {\bf
  27}, 3599 (1994).

\bibitem{Nienhuis96} J. de Gier, B. Nienhuis, Exact Solution of an
  Octagonal Random Tiling Model, {\em Phys. Rev. Lett.} {\bf 76}, 2918
  (1996).

\bibitem{Nienhuis98} J. de Gier, B. Nienhuis, Bethe Ansatz Solution of
  a Decagonal Rectangle-Triangle Random Tiling, {\em J. Phys. A:
  Math. Gen.} {\bf 31}, 2141 (1998).

\bibitem{Widom98} M. Widom, N. Destainville, R. Mosseri, F. Bailly,
Two-Dimensional Random Tilings of Large Codimension, {\sl in}
{\em Proceedings of the 6th International Conference on
Quasicrystals}, (World Scientific, 1998).

\bibitem{Mosseri93B} R. Mosseri, F. Bailly, Configurational entropy
in octagonal tiling models, {\em Int. J. Mod. Phys. B}, Vol 7, {\bf 6}\&{\bf
7}, 1427 (1993).

\bibitem{Elser} V. Elser, Comment on ``Quasicrystals: a new class of
ordered structures'', {\em Phys. Rev. Lett.} {\bf 54}, 1730 (1985).

\bibitem{Duneau} M. Duneau, A. Katz, Quasiperiodic patterns, {\em
    Phys. Rev. Lett.} {\bf 54}, 2688 (1985).

\bibitem{Kalugin} A.P. Kalugin, A.Y. Kitaev, L.S. Levitov, 
Al$_{0.86}$Mn$_{0.14}$: a six-dimensional crystal, {\em JETP Lett.} {\bf
41}, 145 (1985); A.P. Kalugin, A.Y. Kitaev, L.S. Levitov, 6-dimensional
properties of Al$_{0.86}$Mn$_{0.14}$, {\em J. Phys. Lett. France} {\bf 46},
L601 (1985).

\bibitem{DeBruijn81} N.G. de Bruijn, Algebraic theory of Penrose's
non-periodic tilings of the plane, {\em
Kon. Nederl. Akad. Wetensch. Proc. Ser. A} {\bf 43}, 84 (1981).

\bibitem{DeBruijn86} N.G. de Bruijn, Dualization of multigrids, 
{\em J. Phys. France} {\bf 47}, C3-9 (1986).

\bibitem{Socolar85} J.E.S. Socolar, P.J. Steinhardt, D. Levine,
Quasicrystals with arbitrary orientational symmetry, {\em Phys. Rev.} 
B {\bf 32}, No 8, 5547 (1985).

\bibitem{Gahler86} F. G\"ahler, J. Rhyner, Equivalence of the
  generalized grid and projection methods for the construction of
  quasiperiodic tilings, {\em J. Phys. A: Math. Gen.} {\bf 19}, 267
  (1986).

\bibitem{Elser84} V. Elser, Solution of the dimer problem on an hexagonal 
lattice with boundary, {\em J. Phys. A: Math. Gen.} {\bf 17}, 1509 
(1984).

\bibitem{Mosseri93} R. Mosseri, F. Bailly, C. Sire, Configurational 
entropy in random tiling models, {\em J. Non-Cryst. 
Solids}, {\bf 153}\&{\bf 154}, 201 (1993).

\bibitem{These} N. Destainville, Ph.D. Thesis: ``Entropie
  configurationnelle des pavages al\'eatoires et des membranes
  dirig\'ees'', {\sl Th\`ese de l'Universit\'e Paris 6} (1997).

\bibitem{Bibi98} N. Destainville, Entropy and boundary conditions in random
rhombus tilings, {\em J. Phys. A: Math. Gen.} {\bf 31}, 6123 (1998). 

\bibitem{Bailey97} G.D. Bailey, Tilings of zonotopes: discriminental
arrangements, oriented matroids and enumeration, {\sl Minnesota
University Thesis} (1997).

\bibitem{Cohn98} H. Cohn, M. Larsen, J. Propp, The shape of a typical
boxed plane partition, {\em New York J. of Math.} {\bf 4}, 137 (1998).

\bibitem{Cohn9?} H. Cohn, R. Kenyon, J. Propp, A variational principle
for domino tilings, {\sl to appear}, {\em J. of AMS}.

\bibitem{Latapy99} M. Latapy, Generalized integer partitions, 
tilings of zonotopes and lattices, {\sl preprint}.

\bibitem{Kenyon93} R. Kenyon, Tilings of polygons with parallelograms,
{\em Algorithmica} {\bf 9}, 382 (1993).

\bibitem{Elnitsky97} S. Elnitsky, Rhombic tilings of polygons and
classes of reduced words in Coxeter groups, {\em J. Combinatorial
Theory A} {\bf 77}, 193 (1997).

\bibitem{Stanley72} R.P. Stanley, Ordered structures and partitions, 
{\em Memoirs of the AMS} {\bf 119} (1972). 

\bibitem{Brock} V. Strehl, Combinatorics of special functions: facets of
Brock's identity, {\em in S\'eries formelles et combinatoire 
alg\'ebrique}, eds. P. Leroux and C. Reutanauer (University of Qu\'ebec,
Montr\'eal, 1992).

\bibitem{Henley97} C.L. Henley, Relaxation time for a dimer covering
with height representation, {\em J. Stat. Phys.} {\bf 89}, 483 (1997).

\bibitem{Randall98} D. Randall and P. Tetali, Analyzing Glauber 
dynamics by comparison of Markov chains, Proceedings of the
3rd Latin American Theoretical Informatics Symposium, {\em Springer
Lecture Notes in Computer Science}, Vol. {\bf 1380}, 292 (1998).

\bibitem{Randall99} M. Luby, D. Randall, A. Sinclair, Markov chain
algorithms for planar lattice structures, {\sl preprint}.

\bibitem{Wilson99} D.B. Wilson, Mixing times of lozenge tiling and
card shuffling Markov chains, {\sl preprint}.

\bibitem{Mosseri95} R. Mosseri, J.-F. Sadoc, Glass-like properties 
in quasicrystals, {\em Proceedings of the 5th International Conference
on Quasicrystals}, Ed. C. Janot, R. Mosseri (World Scientific, 1995),
p.~747. 

\bibitem{Leuzzi99} L. Leuzzi, G. Parisi, A Tiling Model for Glassy Systems,
{\sl preprint} ({\tt cond-mat/9911020}).

\bibitem{Strandburg90} K.J. Strandburg, P.R. Dressel, Thermodynamic 
behavior of a Penrose-tiling quasicrystal, {\em Phys. Rev.} B {\bf 41},
2469 (1990).

\bibitem{Ishii95} Y. Ishii, Dynamics of phason relaxation on Penrose
lattices, {\em Proceedings of the 5th International Conference on
Quasicrystals}, Ed. C. Janot, R. Mosseri (World Scientific, 1995),
p.~359.

\bibitem{Stanley84} R.P. Stanley, On the number of reduced decompositions of
elements of Coxeter groups, {\em European J. of Comb.} {\bf 5},
359 (1984).

\bibitem{Edelman87} P. Edelman, C. Greene, Balanced tableaux, 
{\em Adv. in Math.} {\bf 63}, 42 (1987).

\bibitem{Sagan} B.E. Sagan, {\em The symmetric group} (Wadsworth and Brooks,
California, 1991).

\bibitem{Knuth92} D.M. Knuth, Axioms and hulls, {\sl in} {\em Lect. Notes 
in Computer Sci.} {\bf 606}, 35 (1992).

\bibitem{Young27} A. Young, On quantitative substitutional analysis, 
{\em Proc. London Math. Soc.}, 2, Vol. {\bf 28}, 255 (1927).

\bibitem{Greene79} C. Greene, A. Nijenhuis et H.S. Wilf, A probabilistic
  proof of a formula for the number of Young tableaux of a given
  shape, {\em Adv. in Math.} {\bf 31}, 104 (1979).

\bibitem{Gessel85} I. Gessel, G. Viennot, Binomial determinants, paths
and hook lenght formulae, {\em Adv. in Math.} {\bf 58}, 300
(1985).

\bibitem{Bjorner93} A. Bj\"orner, M. Las Vergnas, B. Sturmfels,
N. White, G.M. Ziegler, {\em Oriented matroids} (Cambridge University
Press, 1993).

\bibitem{Sturmfels93} B. Sturmfels, G.M. Ziegler, Extention spaces of
oriented matroids, {\em Discrete \& Computational Geom.} {\bf 10}, 23
(1993).


\end{thebibliography}
\end{document}